\documentclass[12pt,a4paper]{article}
\usepackage{amsmath,amssymb,amsthm,amsfonts}
\usepackage[dvips]{graphicx}
\usepackage{epsfig}
\usepackage{siunitx}	
\usepackage{caption}
\usepackage{subcaption}
\usepackage{float}
\usepackage{verbatim}
\usepackage[font=small]{caption}
\usepackage{soul}
\usepackage{physics}
\usepackage{multirow}
\usepackage{url}
\usepackage{enumerate} 
\usepackage{color}
\usepackage{xcolor}

\def\e{\epsilon}

\def\p{\partial}

\def\({\text{\huge (}}
\def\){\text{\huge )}}
\def\bf{\textbf}
\def\]{\text{\huge ]}}
\def\[{\text{\huge [}}

\def\bf{\textbf}

\def\gx{\nabla_{\vb{x}}}
\def\gy{\nabla_{\vb{y}}}

\newcommand{\bi}{\begin{itemize}}
\newcommand{\ei}{\end{itemize}}
\newcommand{\be}{\begin{equation}}
\newcommand{\ee}{\end{equation}}
\newcommand{\ba}{\begin{align}}
\newcommand{\ea}{\end{align}}

\newcommand\nc{\newcommand}
\nc\pad[2]{\frac{\p #1}{\p #2}} \nc\padd[2]{\frac{\p^2 #1}{\p
{#2}^2}} \nc\nd[2]{\frac{d #1}{d #2}} \nc\pat[2]{\frac{D #1}{D
#2}} \nc\ov{\overline} \nc\degree{^{\circ}} \nc\ord[1]{{\cal
O}(#1)} \nc\ra{\rightarrow} \nc\Ra{\Rightarrow} \nc\dint{{\mbox ~
d}}

\DeclareMathOperator{\Pe}{Pe^{-1}}       %%% Peclet number
\DeclareMathOperator{\Da}{Da}		%%% Damkohler number
\newcommand{\bea}{\begin{eqnarray}}
\newcommand{\eea}{\end{eqnarray}}
\newcommand{\beas}{\begin{eqnarray*}}
\newcommand{\eeas}{\end{eqnarray*}}

\title{Assessment of averaged 1D models for column adsorption with 3D computational experiments}
\author{M. Aguareles$^1$ \& F. Font$^2$\\
\small{$^1$Dept. Computer Science, Applied Mathematics and Statistics, Universitat de Girona, Spain} \\
\small{$^2$Dept. Fluid Mechanics, Universitat Politècnica de Catalunya, Spain}
}
\begin{document}
\maketitle

% TODO Reduce abstract: DONE 
\begin{abstract}
In the present manuscript, we formulate a 3D mathematical model describing the capture of a contaminant in an adsorption column. The novelty of our approach involves the description of mass transfer by adsorption via a nonlinear evolution equation defined on the surface of the porous media, while Stokes flow and an advection-diffusion equation describe the contaminant transport through the interstices. Simulations of 3D models with varying microstructures but identical porosity reveal a minimal impact of the microstructure on contaminant distribution within the column, particularly in the radial direction. Using homogenization theory and a periodic microstructure, we rigorously derive a 1D adsorption model that matches the standard form that is found in the literature, but contains two effective coefficients, the dispersion and the permeability, that explicitly incorporate microstructural details of the porous medium. 
The 1D model closely reproduces 3D results, the 1D concentration profiles closely match the cross-section averaged 3D profiles, and the outlet breakthrough curves are nearly identical. We also demonstrate how the 3D simulations converge to the solution of the 1D model as the microstructure is refined. Consequently, our model offers a theoretical foundation for the widely used 1D model, confirming its reliability for investigating, optimizing, and aiding in the design of column adsorption processes for practical applications. 
\end{abstract}

\section{Introduction}

\begin{figure}
\begin{center}
\includegraphics[width=\textwidth]{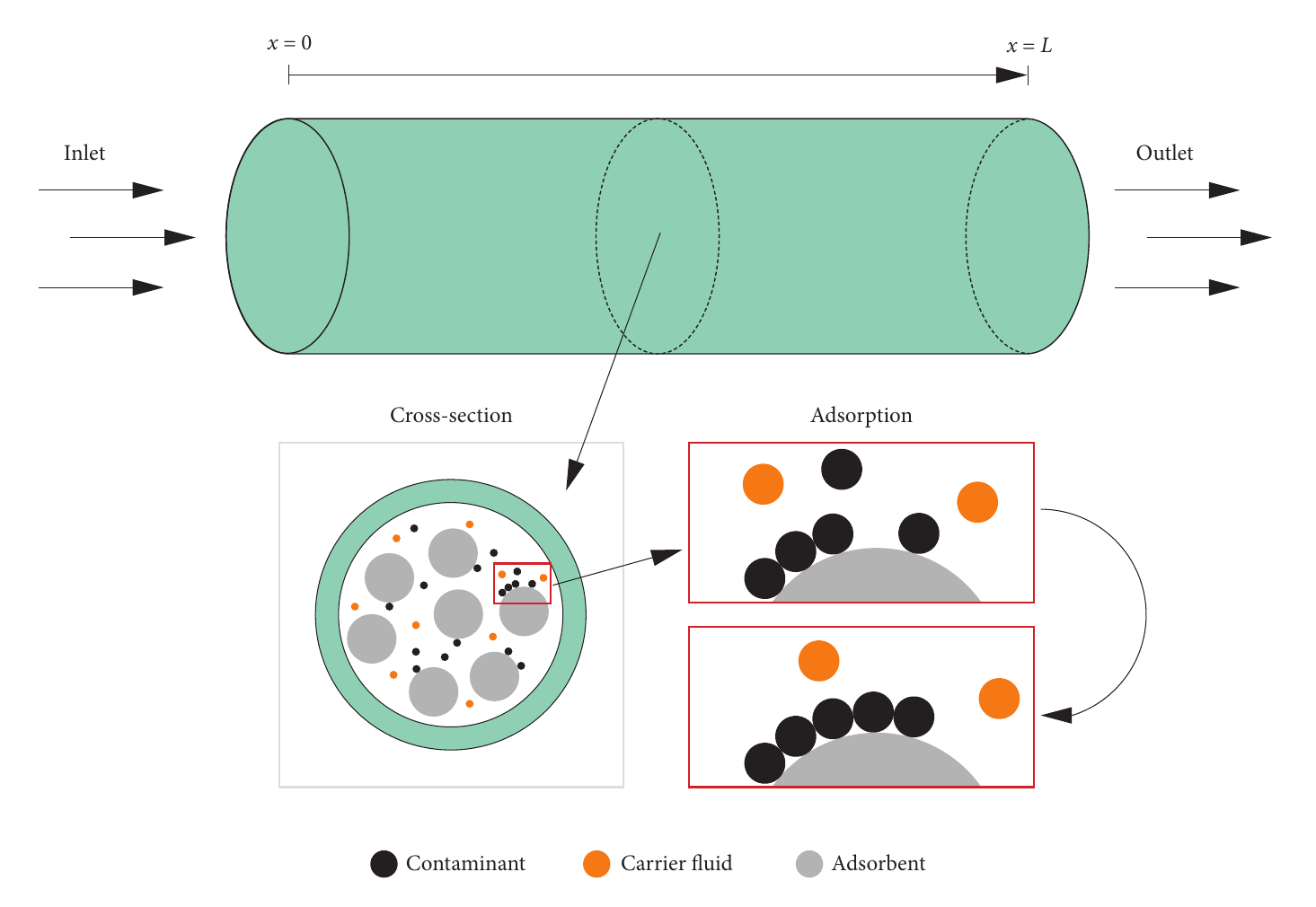}
\caption{Sketch of an adsorption column. The contaminated fluid enters the column through the inlet (at $x=0$ in the illustration). As the fluid flows toward the outlet, contaminant molecules adhere to the surfaces of the adsorbent material. Clean fluid is discharged at the outlet until the adsorbent reaches its saturation point.}
\label{fig:sketch}
\end{center}
\end{figure}

Adsorption columns play a crucial role in mitigating environmental pollutants, including greenhouse gases, volatile organic compounds and emerging contaminants \cite{karimi23,Hong21,ahmed18}. Column sorption involves directing a fluid through a tube packed with an adsorbent material that can selectively capture specific components from the fluid through chemical or physical interactions \cite{Gabelman2017} (see illustration in Fig.~\ref{fig:sketch}). They are widely used in various applications such as water treatment, biogas purification, and the production of biopharmaceuticals \cite{Patel2019,Meena2024,Chase94}. These systems are relatively simple to integrate into industrial processes and can be applied to both liquid and gas-phase contaminants, making them an essential technology for environmental remediation. Despite the environmental benefits, the higher costs associated with current capture technologies diminish their economic appeal. Mathematical models can help address this challenge by providing insights into adsorption kinetics, mass transfer dynamics and column design parameters, enabling more efficient system optimization.

Existing mathematical models for adsorption often focus on highly idealized scenarios and many widely accepted solutions deviate from the model’s foundational assumptions, exhibit unrealistic parameter dependencies, and yield predictions that can be significantly inaccurate (see for instance \cite{BA1920} and \cite{Myers23}). Recently, several publications  have made substantial progress in rigorously establishing robust mathematical models based on the physio-chemistry of adsorption processes (see \cite{Mondal19, Myers20a, Myers20b, SipsPaper, AUTON2024827, Valverde24}, among others). One important contribution in most of these works is the derivation of approximate analytical expressions for the breakthrough curve, i.e. the evolution of the mean concentration at the outlet. 
Deriving these analytical expressions requires formulating and solving a 1D model for the cross-sectional average concentration and adsorbed fraction of the contaminant, $c(x,t)$ and $q(x,t)$, in the column. These expressions are particularly useful for inferring system parameters by fitting them to experimental data.

For incompressible flows with low contaminant concentrations where mass loss due to adsorption may be neglected the one-dimensional model for column adsorption is found to be:
\begin{subequations}
\label{eq:1d}
\begin{alignat}{2}
	&\pdv{c}{t} + u\pdv{c}{x} = D\pdv[2]{c}{x}-\frac{\rho_b}{\phi}\pdv{q}{t}\, ,\quad && x\in(0,L)\,, t\in(0,\infty),\label{eq:c1d}\\
	&\pdv{q}{t} = k_\textrm{ad} c^m(q_\textrm{max}-q)^n-k_\textrm{de} q^n\, , \quad && x\in(0,L)\,, t\in(0,\infty),\\
	& c(x,0)=q(x,0) = 0\, , && x\in(0,L),\\
	& c_\textrm{in}u(t,0^-)  = \left(u c- D\pdv{c}{x}\right)_{x=0^+}\, ,\quad \left. \pdv{c}{x}\right|_{x=L^-}=0\,, \quad && t>0.
	\end{alignat}
	\end{subequations}
where $u(t,x)$ is the intersticial velocity, $D$ is the dispersion coefficient, $\rho_b$ is the density of the bulk material (defined as the ratio of the mass of adsorbent to the column volume), $m,n$ are the global orders of the adsorption reaction, $k_\textrm{ad},k_\textrm{de}$ are the adsorption and desorption rates and $q_\textrm{max}$ is the maximum adsorbed fraction that the adsorbent material can achieve. See \cite{SipsPaper} for details on a derivation of system \eqref{eq:1d} and its solutions in the form of traveling waves. 
Note that $D$ accounts for molecular diffusion as well as axial dispersion caused by microstructural obstacles encountered by contaminant molecules along their path. Theoretical prediction of dispersion effects due to the porous media microstructure is generally complex. Homogenization theory can be used to obtain theoretical estimates of dispersion coefficients and permeabilities \cite{bruna15,ANDERSON201842}, though its application is limited to relatively simple porous media structures. In column adsorption models, the value of $D$ is typically determined through experimental correlations \cite{Delgado2006,Shaf15}.

% TODO Tim's ? hopefully answered: DONE
We begin by formulating a system of partial differential equations for the contaminant concentration in the fluid region of a general 3D filter configuration, explicitly determining the boundary conditions at the surfaces where the adsorption reaction occurs. Surface reactions are inherently complex and involve several phenomenological parameters, such as adsorption and desorption rates, whose direct measurement is virtually impossible. Existing column adsorption models provide expressions for these reactions that are valid only from an averaged perspective, incorporating them as a lumped sink term in the governing equation for contaminant transport (see, for instance, \cite{Zafanelli22,chen19}). Thus, our first contribution is to develop a three-dimensional model based on partial differential equations that allows for the explicit inclusion of adsorption reactions on the surface of the porous media.

A detailed 3D computational model, like the one developed in this study, is valuable for analyzing the impact of the microstructure on the adsorption process. It can be applied to arbitrary geometries, such as those from computed tomography scans of real columns, to assess its effects in realistic settings. However, the simplicity of 1D models like \eqref{eq:1d}, which allow for approximate analytical solutions of breakthrough curves, remains highly valuable to experimentalists. In \cite{SipsPaper}, the model \eqref{eq:1d} was derived using a cross-sectional averaging approach, under the assumption that the porosity of the cross-section does not vary along the axial direction. This is a rather restrictive assumption, which we also address in this paper by instead assuming a constant volumetric porosity. In particular, we employ homogenization techniques to derive equations on a simpler homogeneous domain, where the microstructure contribution explicitly appears in the governing equations. 
We then show how the resulting model can be systematically reduced to the one-dimensional form \eqref{eq:1d}. A key contribution of this reduced model is its explicit incorporation of microstructural effects, captured through the dispersion coefficient and the permeability of the porous medium. In particular, we derive an analytical expression for the permeability as a function of particle diameter and porosity, allowing for the direct computation of flow velocity using Darcy's law. Finally, we compare the solutions of the detailed 3D model with those of the reduced 1D model and demonstrate excellent agreement between them.
%We continue and demonstrate how the resulting model can be reduced to \eqref{eq:1d}, thereby validating the results in \cite{SipsPaper} for a more general configuration. Another key contribution of the derived 1D model is its explicit representation of microstructural effects, expressed through the dispersion coefficient and the permeability of the porous medium. Notably, we derive an explicit formula for the permeability as a function of particle diameter and porosity, enabling the calculation of flow velocity via Darcy's law. Furthermore, we compare the solutions obtained from the detailed 3D model with those from the reduced 1D version and show that they agree remarkably well. This agreement confirms both the robustness of the results in \cite{SipsPaper} and their suitability for data fitting and predicting filter performance. 

%\maria{REPASSAR AIXO quan estigui l'estructura nova:
This article is structured as follows. In Section~\ref{sec:model}, we begin with the derivation and detailed description of a 3D model that captures the evolution of contaminant concentration within the porous media of the column, accounting for adsorption occurring at the adsorbent surfaces. In Section~\ref{sec:hom}, we apply asymptotic homogenization techniques to simplify the original 3D model. The effects of the complex microstructure are reflected in the dispersion coefficient of the reaction-diffusion equation and in the permeability appearing in the flow equations, which reduce to Darcy's law. Additionally, the adsorption boundary condition is incorporated as a sink term in the reaction-diffusion equation. This new model, defined over the column's domain, represents a significant simplification of the original formulation. In Section~\ref{sec:bcs_and_reduced_1D}, the homogenised 3D model is reduced to a one-dimensional form and an explicit expression for the media permeability is provided. In Section~\ref{sec:asses}, we conduct numerical simulations to compare the results of the complete 3D model with the simplified 1D model, showing excellent agreement in the concentration profiles and breakthrough curves produced by both approaches. By assuming the same microstructure as that used in the homogenization, we also demonstrate the convergence of the computational results toward the solution of the simplified 1D model. Finally, Section~\ref{sec:conc} summarizes the conclusions of this work.

%We further demonstrate that the solutions of this system depend solely on the axial direction, leading to equations that are equivalent to those in \eqref{eq:1d}. 

%}

\section{A 3D mathematical model for column adsorption}
\label{sec:model}
\begin{figure}[htb]
\begin{center}
\includegraphics[width=\textwidth]{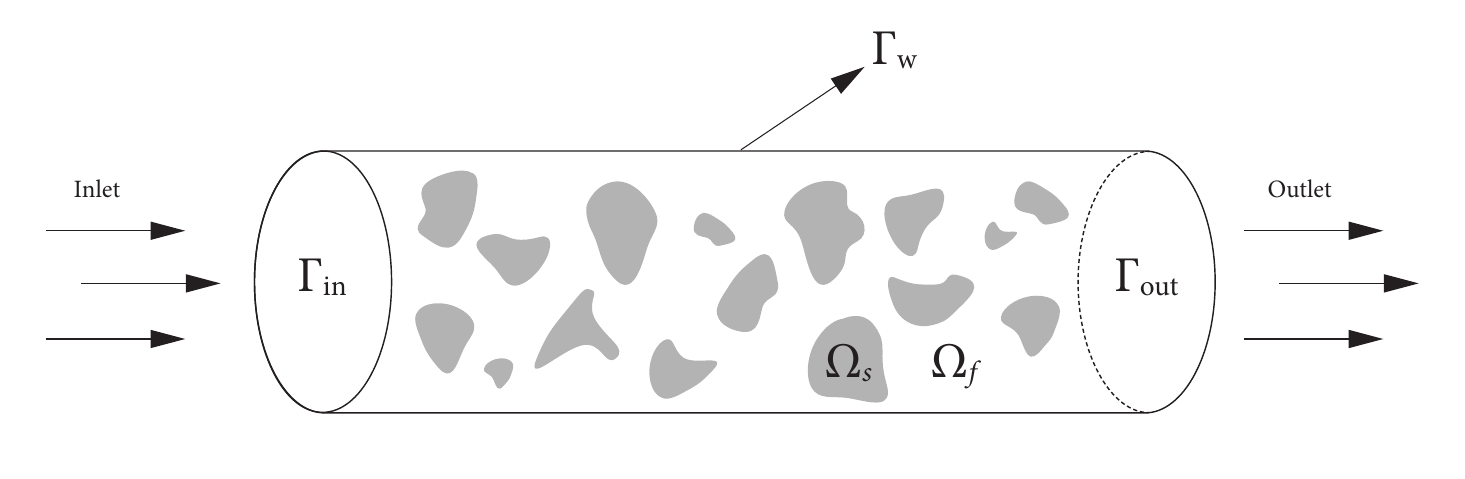}
\end{center}
	\caption{Sketch of an adsorption column with adsorbent particles of different sizes and shapes and the domain's description.}
	\label{fig:gloria}
\end{figure}

Typically, adsorption columns are cylinders with the adsorbent material, usually manufactured as pellets, randomly distributed throughout its interior (see Figure~\ref{fig:gloria}). However, in experimental settings the adsorbent material is smashed and sifted into a powder of grains of a controlled size.  In this section, we present a mathematical model for an adsorption column filled with impermeable particles of arbitrary size and shape. Contaminated fluid is injected at the inlet, and as it flows through the reactor, the contaminant molecules are adsorbed onto the particles' surfaces, resulting in the release of clean fluid at the outlet. We assume that the contaminant concentration is sufficiently low to remain far from saturation, allowing us to neglect any mass loss effects. Additionally, and as explained in \cite{Myers20a}, in most settings the process can be assumed to be isothermal and thus we neglect any heat transfer effects.
	
The interior domain of the column, $\Omega$, is composed of two parts: a solid region occupied by the impermeable solid grains, denoted by $\Omega_s$, and the surrounding space through which the fluid flows, denoted by $\Omega_f$. The porosity is the void fraction of the column, that is 
$$\phi=\frac{|\Omega_f|}{|\Omega_f|+|\Omega_s|}.$$ 
The equation for the concentration of contaminant in the fluid region, $c(\vb{x},t)$ (moles of contaminant per unit volume), corresponds to a standard advection-diffusion equation,
\begin{subequations}
\label{eq:model_ad}
\begin{equation}
\div\left({\cal D}\grad c-\vb{u} c\right)=\pdv{c}{t}\, ,\quad \vb{x}\in\Omega_f\, ,\label{eq:MB}	
\end{equation}
where ${\cal D}$ is the Brownian diffusion and $\vb{u}(\vb{x},t)$ is the flow velocity. In this work, we assume the dilute approximation, under which the diffusion term in \eqref{eq:model_ad} remains linear and independent of solute concentration. This implies that molecular diffusion is governed by Fick's law, with a constant diffusion coefficient. The adsorption reaction, which takes place at the grain's surfaces, is introduced in the model as follows. Denoting by $\vb{n}_s$ the outward normal vector at the solid grains, the flux of contaminant entering the solid region is 
$$-\vb{n}_\textrm{s}\cdot\left({\cal D}\grad c-\vb{u} c\right)|\partial\Omega_s|.$$
This must be equal to the concentration of contaminant (in moles of contaminant per unit volume) retained at the solid region, $|\Omega_s| \pdv*{c^\textrm{ad}}{t}$. It is common practice to express the adsorption rate in terms of the nondimensional contaminant density given by the ratio of the adsorbed mass over the mass of the bulk adsorbent material $m^\text{ad}/m_b$. This magnitude can be averaged over a cross section (as done in \cite{SipsPaper}) or per unit volume to relate it with $c^\text{ad}$, which motivates the following definition for the adsorbed fraction of contaminant
\begin{equation}
\label{def:q}
q(\vb{x},t)=\frac{|\Omega_s|c^\textrm{ad}(\vb{x},t)}{\rho_{b}(|\Omega_s|+|\Omega_f|)}=\frac{(1-\phi)c^\textrm{ad}(\vb{x},t)}{\rho_b}\, ,    \end{equation}
where $\rho_{b}$ is the density of the bulk adsorbent material (defined as the ratio of the mass of adsorbent to the column volume), which is assumed to be constant along the column. Therefore, the boundary condition at the surface of the solid grains can be written as
\begin{equation}
 -\vb{n}_\textrm{s}\cdot\left({\cal D}\grad c-\vb{u} c\right)|\partial\Omega_s| = |\Omega_s|\frac{\rho_b}{1-\phi} \pdv{q}{t},\qquad \vb{x}\in \partial \Omega_s\,.\label{eq:adsorb}
\end{equation}
Equation \eqref{eq:adsorb} must be coupled to a dynamical condition for the adsorption rate at the grain's surface. Following the work in \cite{Sips_1948} we consider the model that is most consistent with the kinetics of the physicochemical reactions that take place in adsorption processes,
\begin{equation}\label{eq:adsorption_vis}
\pdv{q}{t} = k_\text{ad}c^m(q_\text{max} - q)^n - k_\text{de} q^n \, ,\quad \vb{x}\in \partial \Omega_s\, ,
\end{equation}
where $k_\textrm{ad}$ and $k_\textrm{de}$ are the adsorption and desorption rates and $q_\textrm{max}$ is the maximum concentration fraction that the adsorbent can retain. In this work, all three parameters are assumed to be constant. 

\end{subequations}
Equations \eqref{eq:model_ad} must be coupled to equations for the dynamics of the fluid, which assuming a Stokes flow with an incompressible fluid read
\begin{subequations}
\label{eq:model_fluid}
\begin{alignat}{2}
-\grad p+\mu\grad^2\vb{u} & = 0, \qquad &&\vb{x}\in\Omega_f,\, \label{eq:stokes}\\
\div\vb{u} &= 0, &&\vb{x}\in\Omega_f,\label{eq:incomp}\\
\vb{u} &= 0, && \vb{x}\in \partial\Omega_s,\label{eq:NS0}
%\maria{\vb{u}}&=u_\text{in}\vb{e}_1,\quad &&\vb{x}\in \Gamma_\text{in}, \label{last}
\end{alignat} 
\end{subequations}
where $\mu$ is the fluid viscosity, equation \eqref{eq:stokes} is the equation for a Stokes flow, equation \eqref{eq:incomp} accounts for the incompressibility of the fluid, and equation \eqref{eq:NS0} corresponds to a non-slip condition at the grain's surfaces. In Section~\ref{sec:asses}, the flow equations \eqref{eq:model_fluid} will be solved numerically. For this purpose, apart from the no-slip condition \eqref{eq:NS0}, the velocity at the inlet will be approximated as a constant axial value, $\boldsymbol{u}|{\Gamma_{in}} = (u_{in}, 0, 0)$, while atmospheric pressure will be imposed at the outlet, $p|{\Gamma_{out}} = p_{atm}$. We note that the flow equations \eqref{eq:model_fluid} can be solved independently of the concentration equations \eqref{eq:model_ad}. 

\subsection{Nondimensional form}
\begin{figure}
	\begin{center}
	\includegraphics[scale=0.5]{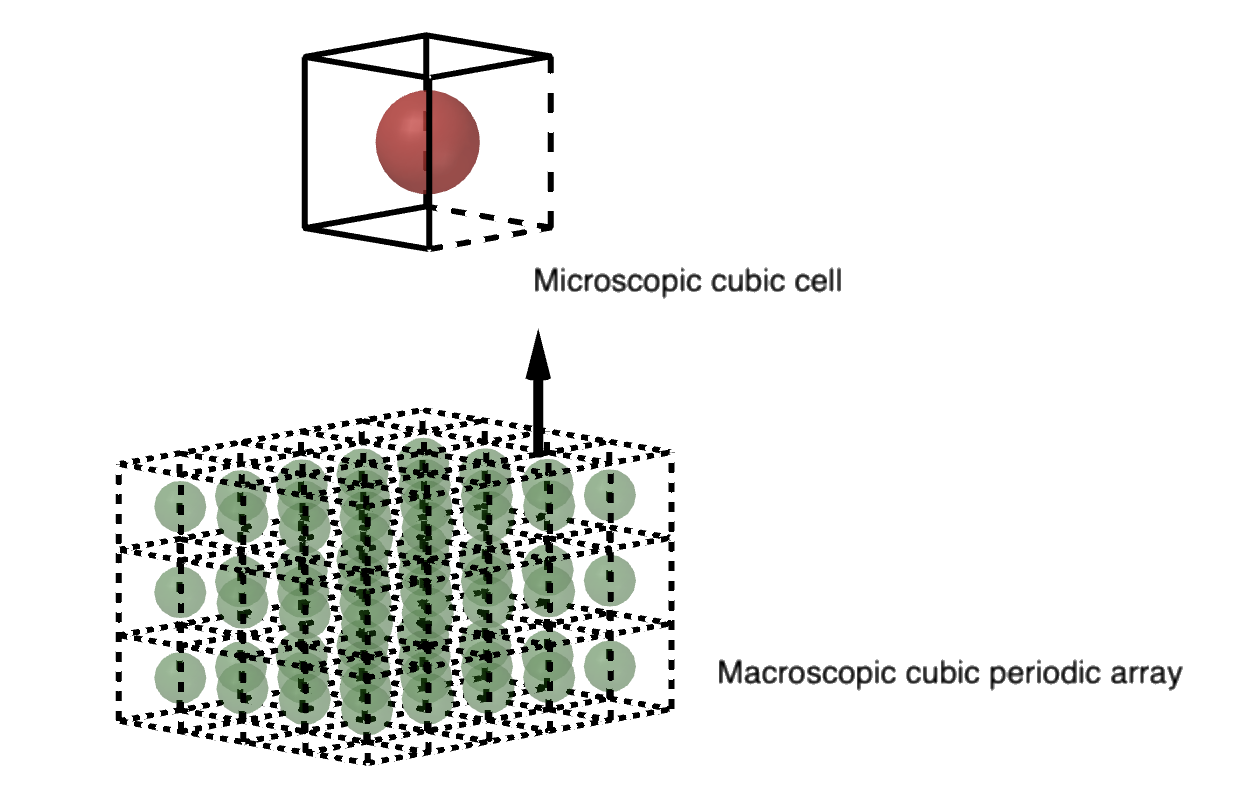}
	\end{center}
	\caption{Grains of adsorbent material in a periodic cubic arrangement. The spheres are separated by distances of size $\ell$. Rescaled microscopic unitary cubic cell.}
	\label{fig:domSpheres}
\end{figure}

In what follows we will assume that the domain is composed of a set of periodically arranged spherical grains whose centres are separated by a distance of size $\ell\ll \mathcal{L}$ (see Figure~\ref{fig:domSpheres}), being $\mathcal{L}$ an adsorption length scale that will be determined later. We also define the small parameter $\e=\ell/\mathcal{L}\ll 1$. We note that $|\Omega_s|$ is the union of a set of spheres with the same radius, and therefore $|\Omega_s|/|\partial\Omega_s|=r/3<\ell/3 =\cal{O}(\e)$ is the inverse of the specific surface area of a spherical grain. 
 %particle that is $r/3$, where $0<r<\ell/2$ is the particles' radius, and it can be written like $r=\nu\ell=\e\mathcal{L}\nu$, where $0<\nu\leq 1/2$.
We also define the following non-dimensional variables 
\begin{equation}
\label{non_dim_Var}
\vb{x} = \mathcal{L} \hat{\vb{x}} ,\quad t=\tau\, \hat{t}\,,\quad c = c_\textrm{in}\, \hat{c}\, , \quad q= q_\textrm{max} \hat{q}\, ,\quad \vb{u} = U\hat{\vb{u}}\, , 
\end{equation}
where $\tau$ is the adsorption time scale that is given by
$$\tau = \frac{1}{k_\text{ad}c_\text{in}^m q_\text{max}^{n-1}}\,,$$
 $\mathcal{L}$ is the corresponding length scale which will be determined later, and $q_\textrm{max}$ is the maximum concentration that the adsorbent can retain. %\maria{Aand $u_\text{in}$ is a constant quantity related with the velocity of the fluid at the inlet. In particular, if the fluid is injected in the column at a constant rate, $u_\text{in}$ would be the constant inlet velocity}. 
 %\cesc{A sensible choice for $U$ is the velocity of the fluid at the column's inlet $u_{\text{in}}$, which is typically a known value in experiments.} 
 Dropping the hats and using the following constants,
$$
\Da =\frac{\mathcal{L}}{U\tau},\quad \Pe = \frac{\mathcal{D}}{\mathcal{L} U},\quad \beta =\frac{|\Omega_s|}{|\partial\Omega_s|\e } \frac{q_\textrm{max}\rho_b  \mathcal{L}}{(1-\phi) c_\textrm{in}U\tau},\quad\delta = \frac{k_\text{de}}{k_\text{ad} c_\text{in}^m q_\text{max}^{n-1}}\,,% \quad p = \frac{u_\textrm{in}\mu}{\e^2 \mathcal{L}}\hat{p}\,, %\red{\tau^{-1} = k_\textrm{ad}c_\textrm{in}^m(c^{\textrm{ad}}_\textrm{max})^{n-1}\, ,\quad \a=\frac{k_\textrm{de}}{k_\textrm{ad}c_\textrm{in}^m}\, ,} 
$$
 yields the non-dimensional version of equations \eqref{eq:model_ad}:
\begin{subequations}
\label{eq:ND_model_ad}
\begin{alignat}{2}
\div\left(\Pe\grad c-\vb{u} c\right)=&\Da\pdv{c}{t}\, ,\quad && \vb{x}\in\Omega_f,\label{eq:ND_MB}\\
\vb{n}_\textrm{s}\cdot\left(\Pe\grad c-\vb{u} c\right) =&-\e\beta\pdv{q}{t},\qquad  && \vb{x}\in \partial\Omega_s,\label{eq:ND_adsorp}\\
 c^m(1 - q)^n - \delta q^n =&\pdv{q}{t} \, ,\qquad && \vb{x}\in \partial \Omega_s\, ,%\\
%\left(\vb{u} c-\Pe\grad c \right) \cdot \vb{n}_\textrm{in} =& \vb{n}_\textrm{in}\cdot\vb{u} ,
%\quad && \vb{x}\in \Gamma_\textrm{in},\label{eq:ND_DC}\\
%\vb{n}_\textrm{out} \cdot \grad c  =& 0,   &&\vb{x}\in \Gamma_\textrm{out}\, ,\label{eq:ND_Out}\\
%\grad c\cdot \vb{n}_\textrm{w}&=0\, , &&\vb{x}\in\Gamma_\textrm{w}
\end{alignat} 
\end{subequations}
where the operator $\grad$ applies to the rescaled variable $\vb{x}$ and the domains are the corresponding rescaled domains. As for the fluid equations, writing
\begin{equation}
\label{eq:p_nondim}
    p = \frac{U\mu}{\e^2 \mathcal{L}}\hat{p}\,,
\end{equation}
and dropping the hats, equations \eqref{eq:model_fluid} read
\begin{subequations}
\label{eq:ND_model_fluid}
\begin{alignat}{2}
-\grad p+\e^2\grad^2\vb{u} =& 0, \quad  &&\vb{x}\in\Omega_f\label{eq:ND_stokes}\\
\div{\vb{u}} =& 0, &&\vb{x}\in\Omega_f.\label{eq:ND_incomp}\\
\vb{u} =& 0, && \vb{x}\in \partial\Omega_s\cup\Gamma_\textrm{w}.\label{eq:ND_NS}
\end{alignat} 
\end{subequations}

In the following section we will exploit the fact that the problem has two separate length scales, i.e. the particles' size and the global tank's dimensions, to derive a set of equations on a simpler (homgenised) domain.

\section{Asymptotic homogenization of the 3D model}
\label{sec:hom}
%In \cite{Myers20a} and \cite{SipsPaper}, the authors perform a cross-sectional average of equation \eqref{eq:MB} and, by taking into account that there is a mass loss through the solid grains' surfaces, an equation for the mean concentration in each section is derived. 
A common approach to formulating one-dimensional adsorption models involves taking a cross-sectional average of equation \eqref{eq:MB}. By accounting for mass loss through the surfaces of the solid grains, this leads to an equation governing the mean concentration within each cross-section (see, for example, \cite{Myers20a,SipsPaper}).
In these derivations the dispersion coefficient ($D$ in equation \eqref{eq:1d}) is not directly connected with the microstructure and it is treated as a fitting parameter. In fact, in most experimental settings, the effective diffusion ends up being small enough to be neglected, (of the order of $10^{-5}$, see for instance \cite{ruthven1984principles,levenspiel1998chemical}). 
%\maria{However, even if the dispersion is found to be small, the key quantity to control is the inverse P\'clet number, which also depends on the velocity and length scale, so it can still exert a non-negligible influence on the solutions.} 
They also assume that the porosity in each cross section, defined as the area occupied by the fluid divided by the total area, is constant,  which significantly limits the possible domain configurations. 

In what follows we will derive a set of equations on a simpler macroscale domain whose properties capture the effect of the actual microscopic geometry. To do so we use the method of multiple scales (MMS) and start by introducing a microscale variable, $\vb{y}:=\vb{x}/\e$, and as is standard in the MMS, assume that the macroscopic variable, $\vb{x}$, and the microscopic one, $\vb{y}$, are independent. Therefore, and abusing notation,  we continue by writing $c(\vb{x},t) = c(\vb{x},\vb{y},t)$, $q(\vb{x},t) = q(\vb{x},\vb{y},t)$, $\vb{u}(\vb{x},t)=\vb{u}(\vb{x},\vb{y},t)$, and $p(\vb{x},t) = p(\vb{x},\vb{y},t)$. The microscale variable,  $\vb{y}$, is defined in a unitary cubic domain, $[-1/2,1/2]^3=\omega_f\cup\omega_s$, being $\omega_s=\{|\vb{y}|\leq r/\ell:=\nu\}$ the particle's domain and $\omega_f$ the surrounding fluid region. We note that $|\omega_f|=\phi$. The boundary of the fluid region in the cubic cell is given by $\partial\omega_f=\{|\vb{y}|= \nu\}\cup\partial([-1/2,1/2]^3)$. The gradient operator in the new variables $(\bf{x},\bf{y})$ reads
$$
\grad = \gx + \frac{1}{\e} \gy \, , 
$$
and so equations \eqref{eq:ND_model_ad} for the contaminant concentration, $c(\vb{x},\vb{y},t)$, and for the adsorbed fraction, $q(\vb{x},\vb{y},t)$,  read
\begin{subequations}
\label{eq:exp_eqns_c}
\begin{alignat}{2}
\Pe\left(\gx^2 c + \frac{1}{\e} \gx\gy c+\frac{1}{\e} \gy\gx c + \frac{1}{\e^2}\gy^2 c\right)&  &&\\
-\gx\cdot(\vb{u} c)- \frac{1}{\e}\gy\cdot(\vb{u} c)&=\,\Da\pdv{c}{t}, &&\bf{y}\in\omega_f\, , \\
    \vb{n}_\textrm{s}(\vb{x})\left(\Pe\gx c + \frac{\Pe}{\e} \gy c -\vb{u} 
    c\right)&=-\e\beta\pdv{q}{t},\quad &&|\vb{y}|= \nu\, ,\\
%\pdv{c^\textrm{ad}}{t}&= \,\Tilde{\mathcal{Q}}(c, c^\textrm{ad}), &&\bf{y}\in\partial\omega_s\, ,\\
 c(\bf{x},\bf{y},t),\,q(\bf{x},\bf{y},t),\,\vb{u}(\bf{x},\bf{y},t),\,\,&  \bf{y}\textrm{-periodic,}&&\bf{y}\in\omega_f\, . \nonumber
\end{alignat} 
\end{subequations}
The flow equations, which can be solved independently of the concentration equations, read
\begin{subequations}
\label{eq:exp_eqns_flow}
\begin{alignat}{2}
-\gx  p-\frac{1}{\e}\gy p+\e^2\gx^2 \vb{u} + \e\gx\gy \vb{u}+\e\gy\gx \vb{u} + \gy^2 \vb{u}=&0, &&\bf{y}\in \omega_f \, ,\\
\gx\cdot \vb{u} + \frac{1}{\e}\gy\cdot \vb{u} =&0 &&\bf{y}\in \omega_f \, ,\label{eq:incompMicro}\\
\vb{u} =& 0, &&|\vb{y}|= \nu\,,\\
p(\bf{x},\bf{y},t),\,\vb{u}(\bf{x},\bf{y},t),\,\,& \bf{y}\textrm{-periodic,}\quad&&\bf{y}\in\omega_f\, .\nonumber
\end{alignat} 
\end{subequations}
We thus proceed and introduce the expansions
\begin{align*}
     p(\vb{x},\vb{y},t),&\sim p_0(\vb{x},\vb{y},t) + \e\, p_1(\vb{x},\vb{y},t) +\cdots,\\
     \vb{u}(\vb{x},\vb{y},t)&\sim \vb{u}_0(\vb{x},\vb{y},t)+\e \vb{u}_1(\vb{x},\vb{y},t)+\cdots,
\end{align*}
as $\e\to 0$ and obtain, at leading order,
\begin{equation}
\gy p_0 = 0,\quad \vb{y}\in\omega_f,\quad \textrm{and $p_0$ is $\vb{y}$ -periodic,}
\end{equation}
whose only possible solution is $p_0(\vb{x},\vb{y},t) = p_0(\vb{x},t)$. The order one terms give,
\begin{subequations}
\label{eq:1stFluid}
\begin{alignat}{2}
-\gy p_1+\gy^2 \vb{u}_0 =\, & \gx p_0, \quad && \vb{y}\in\omega_f,\\
\gy\cdot \vb{u}_0 =\, & 0, \quad && \vb{y}\in\omega_f,\\
\vb{u}_0 =\, &  0,\quad && |\vb{y}|= \nu,\\
 \vb{u}_0(\bf{x},\bf{y},t),\, p_1(\bf{x},\bf{y},t) \quad  & \textrm{are $\vb{y}$ -periodic,} && \nonumber
\end{alignat}    
\end{subequations}
whose solution can be written like
\begin{subequations}
        \begin{align}
    \vb{u}_0(\vb{x},\vb{y},t) &= -\sum_{j=1}^3\left(\vb{K}_j(\vb{y})\pdv{p_0(\vb{x},t)}{x_j}\right)\, ,\label{eq:u0}\\
    p_1(\vb{x},\vb{y},t) &= -\sum_{j=1}^3\left(\Pi_j(\vb{y})\pdv{p_0(\vb{x},t)}{x_j}\right) + \tilde{p}(\vb{x},t)\, ,
\end{align}
\end{subequations}
where $\tilde{p}(\vb{x},t)$ is an arbitrary function independent of $\vb{y}$ which will not play any role in the leading order approximation. Denoting by $\delta_{ij}$ the Delta Kronecker function ($\delta_{ij}=1$ if $j=i$ and 0 otherwise), $\Pi_j, \vb{K}_j = (K_{j1}, K_{j2}, K_{j3})$ are the solutions of
\begin{subequations}
\label{eq:cell_up}
\begin{alignat}{2}
    \delta_{ij}-\pdv{\Pi_i}{y_j}+\gy^2 K_{ij} = \, & 0,\qquad && \vb{y}\in\omega_f, \\
     \pdv{K_{j1}}{y_1}+\pdv{K_{j2}}{y_2}+\pdv{K_{j3}}{y_3} = \, & 0,\qquad && \vb{y}\in\omega_f, \\
    K_{ji}=\, & 0, \qquad && |\vb{y}|= \nu, 
\end{alignat}    
and
\begin{equation}
 \Pi_{j}, K_{ij}\quad  \vb{y}\textrm{-periodic, and}  \int_{\omega_f}\Pi_{j}(\vb{y})  \dd y = \int_{\omega_f} K_{ij}(\vb{y}) \dd y =0, 
\end{equation}
\end{subequations}
for $i,j=1..3$, which have unique solutions. The (constant) interstitial velocity term in equation \eqref{eq:1d} corresponds to the intrinsic average given by
\begin{equation}
\label{eq:u_int}
	\vb{u} (\vb{x},t) = \frac{1}{|\omega_f|}\int_{\omega_f} \vb{u}_0(\vb{x},\vb{y},t)\, \dd \vb{y}= \frac{1}{\phi}\int_{\omega_f} \vb{u}_0(\vb{x},\vb{y},t)\, \dd \vb{y}\, \, .
\end{equation}
Integrating \eqref{eq:u0} over the fluid cell domain,
\begin{equation}
\label{eq:u_int_2}
 \int_{\omega_f} \vb{u}_0(\vb{x},\vb{y},t)\, \dd \vb{y} = -\mathcal{K} \,\gx p_0(\vb{x},t)\, ,
\end{equation}
where $\mathcal{K} =(k_{ij})$ is the symmetric 3 dimensional matrix for the permeability, whose entries are given by
$$
k_{ij}= \int_{\omega_f} K_{ij}(\vb{x},\vb{y})\, \dd \vb{y}\, .
$$
In this particular geometry with spherical grains $\mathcal{K}$ is a diagonal matrix and, in fact, the coefficients $k_{ii}=k$ for all $i=1,2,3$. 

Defining the intrinsic pressure average like
$$p(\vb{x},t) =\frac{1}{|\omega_f|}\int_{\omega_f} p_0(\vb{x},t)\,\dd\vb{y} = \frac{1}{\phi}\int_{\omega_f} p_0(\vb{x},t)\,\dd\vb{y} = p_0(\vb{x},t)\,,$$
and using the intrinsic average for the velocity defined in \eqref{eq:u_int}, expression \eqref{eq:u_int_2} becomes the well-known Darcy's law, which for spherical grains reads
\begin{equation}
\label{eq:darcy}
\vb{u} (\vb{x},t) = -\frac{k}{\phi} \,\gx p(\vb{x},t)	\,,
\end{equation}
for the macroscopic model. We now note that the following order for the incompressibility equation \eqref{eq:incompMicro} is given by
\begin{equation}
\label{eq:vel1}
	\gx\cdot \vb{u}_0+\gy\cdot \vb{u}_1 =\,  0, \quad \vb{y}\in\omega_f.
\end{equation}
An incompressibility condition for the average velocity, $\vb{u}(\vb{x},t)$, is found by integrating \eqref{eq:vel1} over the cell fluid domain,
\begin{equation*}
	\gx\cdot \int_{\omega_f}\vb{u}_0\,\dd\vb{y}+\int_{\omega_f}\gy\cdot \vb{u}_1 \,\dd\vb{y} =\,  0, \quad \vb{y}\in\omega_f.
\end{equation*}
The Divergence Theorem on the second integral, along with the non-slip condition at the grain's surface and the $\vb{y}$-periodicity of $\vb{u}_1$ gives
$$ \int_{\omega_f}\gy\cdot \vb{u}_1 \,\dd\vb{y}=0\,.$$
Therefore, using the definition of the intrinsic average for the velocity provided in \eqref{eq:u_int}, one obtains the incompressibility condition for the intrinsic average velocity,
\begin{equation}
	\label{eq:incompU}
	\gx\cdot \vb{u}=\,  0, \quad \vb{y}\in\Omega.
\end{equation}
where $\Omega = \Omega_s\cup\Omega_f$, that is to say, it is the macroscopic (dimensionless) domain corresponding to the whole column. 

As for the concentration problem, we do as before and we introduce the expansion $c(\vb{x},\vb{y},t),\sim c_0(\vb{x},\vb{y},t) + \e\, c_1(\vb{x},\vb{y},t) + \e^2\, c_2(\vb{x},\vb{y},t) +\cdots$,
%\begin{align*}
%    c(\vb{x},\vb{y},t),&\sim c_0(\vb{x},\vb{y},t) + \e\, c_1(\vb{x},\vb{y},t) + \e^2\, c_2(\vb{x},\vb{y},t) +\cdots,\\
%     c^\textrm{ad}(\vb{x},\vb{y},t),&\sim c^\textrm{ad}_0(\vb{x},\vb{y},t) + \e\, c^\textrm{ad}_1(\vb{x},\vb{y},t) + \e^2\, c^\textrm{ad}_2(\vb{x},\vb{y},t) +\cdots,\\
%\end{align*}
which substituting in \eqref{eq:exp_eqns_c} gives, to leading order,
\begin{subequations}
\begin{alignat}{2}
    \gy^2 c_0 =\,& 0,\quad &&\vb{y}\in\omega_f\\
     \vb{n}_\textrm{s}(\vb{x})\cdot \gy c_0=\,&0, \quad &&|\vb{y}|= \nu\\
    c_0(x,y,t)\quad  \vb{y}&\textrm{-periodic,}\nonumber
\end{alignat}
\end{subequations}
whose solution is independent of the microscopic variable $\vb{y}$, that is $c_0(\vb{x},\vb{y},t) = c_0(\vb{x},t)$. Continuing to the following order, the first order approximation of the concentration is found to satisfy
\begin{subequations}
\label{eq:1st}
\begin{alignat}{2}
      \gy^2 c_1 = \, & 0,\qquad && \vb{y}\in\omega_f \\
    \vb{n}_\textrm{s}(\vb{x})\cdot\gy c_1= \, &-\vb{n}_\textrm{s}(\vb{x})\cdot \gx c_0, \qquad &&|\vb{y}|= \nu\\
    c_1(x,y,t) & \quad \textrm{is $\vb{y}$-periodic.} && \nonumber
\end{alignat}   
\end{subequations}
To solve \eqref{eq:1st} we note that:
$$
\gx c_0(\vb{x},t) = \sum_{j=1}^3\left(\vb{e}_j\pdv{c_0(\vb{x},t)}{x_j}\right),
$$
so one can write
    \begin{equation}
    \label{sol:c1}
    c_1(\vb{x},\vb{y},t) = \tilde{c}(\vb{x},t) -\sum_{j=1}^3\left(\xi_j(\vb{x},\vb{y})\pdv{c_0(\vb{x},t)}{x_j}\right),
\end{equation}
where $\tilde{c}(\vb{x},t)$ is an arbitrary function independent of $\vb{y}$ that will not play any role in the leading order approximation, and where the functions $\xi_j(\vb{x},\vb{y})$ satisfy the following cell problem:
\begin{subequations}
\label{eq:cell_c}
\begin{alignat}{2}
    \gy^2 \xi_j &= \,  0,\qquad && \vb{y}\in\omega_f\, , \\
     \vb{n}_\textrm{s}(\vb{x})\gy \xi_j& =\,  \frac{y_j}{\nu}, \qquad &&  |\vb{y}|= \nu\, ,\\
    &\textrm{$\xi_j$ is $\vb{y}$-periodic and} && \int_{\omega_f}\xi_j(\vb{x},\vb{y})\, \dd \vb{y} = 0.   
\end{alignat}    
\end{subequations}
At this point we would like to remark the fact that both cell problems, \eqref{eq:cell_up} and \eqref{eq:cell_c}, could be written for any other cell geometry, so the following derivation would also be valid for any other grain's and/or cell's shape. In the particular case of a cubic arrangement, the porosity is bounded by the fact that the maximum particle's radius is 1/2, and so consistent porosities in this arrangement satisfy $\phi>1-\pi/6\approx 0.48$.

Continuing to the next order, the equation for the concentration yields
\begin{subequations}
\label{eq:2nd}
\begin{alignat}{2}
    \Pe&\left(\gx^2 c_0+2\gx\gy c_1+\gy^2 c_2\right)-\gx(\vb{u}_0c_0)-\gy(\vb{u}_0c_1+\vb{u}_1c_0) =\, \Da\pdv{c_0}{t}, \,\, &&\vb{y}\in\omega_f\, , \\
     \vb{n}_\textrm{s}&\left(\Pe(\gx c_1+\gy c_2)- \vb{u}_0c_1-\vb{u}_1c_0\right) =\,-\beta\pdv{q}{t} , &&|\vb{y}|= \nu\, ,
\end{alignat}    
\end{subequations}
and $c_2$ is $\vb{y}$ -periodic. We now integrate equations \eqref{eq:2nd} over the cell fluid region, $\omega_f$, which gives:
\begin{equation*}
\begin{split}
&    \gx\cdot\left[\int_{\omega_f}\left(\Pe\left(\gx c_0+\gy c_1\right)-\vb{u}_0 c_0\right)\, \dd \vb{y} \right]\\
&\hspace{1cm}+\int_{\omega_f}\gy\cdot\left(\Pe(\gx c_1+\gy c_2)-(\vb{u}_0c_1+\vb{u}_1 c_0)\right)\,\dd \vb{y} = \Da\int_{\omega_f}\pdv{c_0}{t}\, \dd \vb{y}\, .
\end{split}
\end{equation*}
Using the Gauss-Green theorem along with the boundary and periodicity conditions the previous equation reduces to
\begin{equation}
\label{eq:int_c2}
\gx\cdot\left[\int_{\omega_f}\left(\Pe(\gx c_0+\gy c_1)-\vb{u}_0c_0\right)\,\dd y\right] -\beta\int_{|\vb{y}|= \nu} \pdv{q}{t} \textrm{d}\,\ell = \Da|\omega_f| \pdv{c_0}{t}.
\end{equation}
Dividing  by $|\omega_f|=\phi$ and using \eqref{eq:u_int} and \eqref{sol:c1}, equation \eqref{eq:int_c2} becomes
\begin{equation*}
%\label{eq:int_c2B}
\gx\cdot\left[\Pe\left(\gx c_0-\sum_{j=1}^3\frac{\partial c_0}{\partial x_j}\frac{1}{\phi}\int_{\omega_f} \gy \omega_j(\vb{x},\vb{y})\,\dd \vb{y}\right)\right]-\gx\cdot(c_0 \vb{u})= \frac{\beta}{\phi}\int_{|\vb{y}|= \nu} \pdv{q}{t} \textrm{d}\ell +\Da \pdv{c_0}{t}.
\end{equation*}
In \cite{SipsPaper}, the authors obtain equations \eqref{eq:1d} in terms of an intrinsic average over each cross section. Since we want to compare our homogenized version with the one obtained in \cite{SipsPaper}, we denote the intrinsic average for the concentration as $c(\vb{x},t)$,
\begin{equation}
	\label{eq:c_int}
	c(\vb{x},t) = \frac{1}{\phi}\int_{\omega_f} c(\vb{x},\vb{y},t)\, \dd \vb{y} \sim \frac{1}{\phi}\int_{\omega_f} c_0(\vb{x},t)\, \dd \vb{y} = c_0(\vb{x},t)\,.
\end{equation}
We note that $q$ is a magnitude that represents the amount of contaminant trapped at each grain, so it is intrinsically independent of $\vb{y}$. Therefore,
$$\frac{\beta}{\phi}\int_{|\vb{y}|= \nu} \pdv{q}{t} \textrm{d}\ell = \frac{\beta|\partial\omega_s|}{\phi}\pdv{q}{t} = \frac{|\Omega_s||\partial\omega_s|}{\e |\partial\Omega_s|} \frac{q_\textrm{max}\rho_b  \mathcal{L}}{(1-\phi) c_\textrm{in}U\phi\tau} \pdv{q}{t}= \frac{q_\textrm{max}\rho_b  \mathcal{L}}{ c_\textrm{in}U\phi\tau}\pdv{q}{t}\,,$$
where we have used that
$$ \frac{|\Omega_s|}{|\partial\Omega_s|} =\frac{|\omega_s|}{|\partial\omega_s|}=\frac{1-\phi}{|\partial\omega_s|}\,.$$
We also recall that $\mathcal{L}$ has not been fixed yet. Choosing 
$$ \mathcal{L} = \frac{ c_\textrm{in}U\phi}{q_\textrm{max}\rho_b}\tau\,,$$ 
and defining the tensor $\mathcal{A}_\text{disp}=(d_{ij}):\Omega\to\mathbb{R}^{3\times 3}$ where, $d_{ij}:\Omega\to\mathbb{R}$, with $i,j=1,\ldots,3$ and
\begin{equation}
\label{eq:dij}
    d_{ij}=\int_{\omega_f}\left(\delta_{ij}-\frac{1}{\phi}\pdv{\xi_j}{y_i}\right)\,\dd \vb{y},
\end{equation}
yields,
\begin{equation}
\gx\cdot\left(\Pe\mathcal{A}_\text{disp}\gx c_0-c_0\bf{u}\right) =\, \Da \pdv{c_0}{t}+ \pdv{q}{t}\, , \quad \vb{x}\in\Omega.
\end{equation}
The tensor $\mathcal{A}_\text{disp}$ carries the information of the microstructure and it can be computed for any cell or grain geometry. In this particular setting where the cells are cubic and the grains are spheres, $\mathcal{A}_\text{disp}$ is a diagonal matrix:
\begin{equation}
\label{eq:A}
    d_{ij} = 0\, , \quad \textrm{if $i\neq j$, and}\quad d_{11}=d_{22}=d_{33}=d\,.
\end{equation}
The exact values of $d$ can be computed using Rayleigh's multipole method (see \cite{Rayleigh92}). In particular, for a cubic lattice with spherical grains,
\begin{equation}
\label{eq:d}
d = \frac{1}{\phi}\left(1-\frac{3(1-\phi)}{3-\phi-0.3914(1-\phi)^{10/3}}\right)\, \quad \text{for}\quad \phi > 1-\pi/6\approx 0.48\, .	
\end{equation}
 We then define an effective inverse P\'eclet number:
$$\text{Pe}_\text{eff}^{-1} = d\Pe = \frac{d\mathcal{D}}{\mathcal{L} U}= \frac{D}{\mathcal{L} U}\, ,$$
where $D$ is the effective dispersion of the media. Therefore, the equation for the intrinsic average concentration defined in \eqref{eq:c_int} at the macroscopic scale is 
\begin{subequations}
\label{eq:averaged}
\begin{alignat}{2}
 \text{Pe}_\text{eff}^{-1}\,\grad^2 c-\div(c\vb{u}) =\,& \Da \pdv{c}{t}+ \pdv{q}{t}\, , \quad && \vb{x}\in\Omega,\\
  c^m(1 - q)^n - \delta q^n =&\pdv{q}{t} , \quad && \vb{x}\in\Omega.
    % \pdv{C^\textrm{ad}}{t} =\, &\Tilde{\mathcal{Q}}(C, C^\textrm{ad})\red{=C^m(1-C^\textrm{ad})^n-\a (C^\textrm{ad})^n}\, ,\quad  && \vb{x}\in\Omega,\\
% \pdv{Q}{t} = & \, C^m(1-Q)^n-\alpha Q^n\,, \quad && \vb{x}\in\Omega,\\
% \red{\div \vb{U} }=\,& 0\, , \quad &&\vb{x}\in\Omega,\\
% \vb{U}(\vb{x},t) = \,& -\mathcal{K} \,\gx p_0(\vb{x},t)\, , && \vb{x}\in\Omega\, ,
\end{alignat}
We note that $\text{Pe}_\text{eff}^{-1}$ now depends on the dispersion coefficient which depends on the Brownian diffusion, $\mathcal{D}$ and on the media porosity, $\phi$. The velocity, $\vb{u}$,  satisfies Darcy's law \eqref{eq:darcy} and the incompressibility condition \eqref{eq:incompU}.
\end{subequations}

\section{Boundary conditions and reduction to a 1D model}\label{sec:bcs_and_reduced_1D}

In Section~\ref{sec:model} we have provided a model for the filter's performance in the interior of the adsorption column. In particular, equations \eqref{eq:model_ad} describe the evolution of the contaminant's concentration in the fluid region, $c(\vb{x},t)$, and the adsorbed fraction of contaminant, $q(\vb{x},t)$, which are coupled to equations \eqref{eq:model_fluid} for the fluid velocity, $\vb{u}(\vb{x},t)$, in the columns interstices. The asymptotic homogenisation approach outlined in Section~\ref{sec:hom} has been employed to derive the following averaged versions of \eqref{eq:model_ad}-\eqref{eq:model_fluid} in the column
\begin{subequations}
\label{eq:average3D}
\begin{alignat}{2}
 \text{Pe}_\text{eff}^{-1}\,\grad^2 c-\div(c\vb{u}) &= \Da \pdv{c}{t}+ \pdv{q}{t}\, , \quad && \vb{x}\in\Omega, \label{eq:3D_concentration}\\
  c^m(1 - q)^n - \delta q^n &=\pdv{q}{t} , \quad && \vb{x}\in\Omega,\\
%\end{alignat}
%\begin{alignat}{2}
\vb{u} &= -\frac{k}{\phi} \,\grad p\,,\quad && \vb{x}\in\Omega\,,\label{eq:Darcy}\\
\grad\cdot \vb{u}  &= 0\, , && \vb{x}\in\Omega\,.\label{eq:incomp3}
\end{alignat}
\end{subequations}

To close the problem  we must now specify appropriate boundary conditions at the filter's boundaries for the governing equations \eqref{eq:average3D}. We assume a Dankwert condition $\left(\vb{u} c-D\grad c \right)\cdot \vb{e}_1 = \vb{e}_1\cdot\vb{u} c_\text{in}$ at the inlet, in dimensional form (see \cite{DANCKWERTS19531}), we impose a no-flux condition $D\grad c \cdot \vb{e}_1= 0$ at the outlet (see \cite{Pear59} for a justification of the convenience of this simplification) and impermeability at the reactor's wall $D\grad c\cdot \vb{n}_\textrm{w}=0$, with $\vb{n}_\text{w}$ being the normal vector to the inner wall of the column and $\vb{e}_1$ the unitary vector in the axial direction. Using the dimensionless variables \eqref{non_dim_Var} and dropping the hat notation, these boundary conditions can be written as
\begin{subequations}
\label{eq:BC_concentration}
\begin{alignat}{2}
\left(\vb{u} c-\text{Pe}_\text{eff}^{-1}\grad c \right)\cdot \vb{e}_1 &= \vb{e}_1\cdot\vb{u} ,  \quad && \vb{x}\in \Gamma_\text{in}\,,\label{eq:DC}\\
\grad c \cdot \vb{e}_1&= 0\,, && \vb{x}\in \Gamma_\text{out}\,,\label{eq:Out}\\
\grad c\cdot \vb{n}_\textrm{w}&=0\, , &&\vb{x}\in\Gamma_\textrm{w}\,. \label{eq:impermeability}
\end{alignat}
\end{subequations}
Note that Darcy's law \eqref{eq:Darcy} does not involve derivatives of the velocity, so typically a pressure drop across the column is prescribed through the boundary conditions
\begin{subequations}
\label{eq:BC_pressure}
\begin{alignat}{2}
p &= p_0 ,  \quad && \vb{x}\in \Gamma_\text{in}\,,\\
p &= p_{\text{atm}}\quad && \vb{x}\in \Gamma_\text{out}\,,
\end{alignat}
\end{subequations}
where $p_{\text{atm}}$ is the dimensionless atmospheric pressure and $p_0>p_{\text{atm}}$.   

\subsection{Reduction to a one-dimensional problem}\label{sec:1D_reduction}

The domain $\Omega$ is now a cylindrical column without internal obstacles, making cylindrical coordinates $(x, r, \varphi)$ a natural choice. Due to radial symmetry, all derivatives with respect to the angular coordinate $\varphi$ in \eqref{eq:average3D} vanish.

We further assume a plug flow profile, meaning the velocity field takes the form $\vb{u} = (u(x), 0, 0)$. Then, applying the incompressibility condition \eqref{eq:incomp3} requires that $u(x)$ is constant. Under this assumption, only the $x$-components in the advection terms of \eqref{eq:3D_concentration} and \eqref{eq:Darcy} remain nonzero. In fact, \eqref{eq:Darcy} can be readily integrated and boundary conditions \eqref{eq:BC_pressure} applied to give
\begin{equation}\label{eq:Darcy_sol}
    u = -\frac{k}{\phi} \frac{(p_{\text{atm}}-p_0)}{L}\,, 
\end{equation}
where $L$ is the dimensionless length of the column.

Additionally, we impose no-flux (impermeability) boundary conditions at the column walls \eqref{eq:impermeability}, and assume the incoming fluid is well-mixed with a uniform concentration. These conditions ensure that the concentration becomes independent of $r$, so that $c(x, r, \varphi) = c(x)$. Under these simplifications, the system \eqref{eq:average3D} reduces to
\begin{subequations}
\label{eq:1D-B}
\begin{alignat}{2}
  \text{Pe}_\text{eff}^{-1}\pdv[2]{c}{x}-u\pdv{c}{x} &=\, \Da \pdv{c}{t}+ \pdv{q}{t}\, , \quad && x\in(0,L),\, t>0,\\
 c^m(1 - q)^n - \delta q^n &=\pdv{q}{t} , \quad && x\in(0,L),\, t>0,
 \end{alignat}
with the constant velocity $u$ described by \eqref{eq:Darcy_sol}. These equations are subject to the boundary conditions 
\begin{alignat}{2} \label{eq:1D-B_bc}
 u = \left(u c- D\pdv{c}{x}\right)_{x=0^+}\, ,\quad \left. \pdv{c}{x}\right|_{x=L^-}=0\,, \quad && t>0\,,
\end{alignat}
\end{subequations}
and the initial conditions $c(x,0)=q(x,0)=0$. Note that, since $u$ is fully determined via \eqref{eq:Darcy_sol}, one can choose the scale $U$ in \eqref{non_dim_Var} to be the dimensional version of \eqref{eq:Darcy_sol} so that $u=1$ in \eqref{eq:1D-B}.

\subsection{Equations in dimensional form}

The dimensionless problem \eqref{eq:1D-B} can be rewritten in dimensional form by reverting the change of variables defined in \eqref{non_dim_Var}, leading to
\begin{subequations}
\label{eq:1d_vis}
\begin{alignat}{2}
&\pdv{c}{t} + u\pdv{c}{x} = d\mathcal{D}\pdv[2]{c}{x}-\frac{\rho_b}{\phi}\pdv{q}{t}\, ,\quad && x\in(0,L)\,, t\in(0,\infty),\label{eq:c1d}\\
&\pdv{q}{t} = k_\textrm{ad} c^m(q_\textrm{max}-q)^n-k_\textrm{de} q^n\, , \quad && x\in(0,L)\,, t\in(0,\infty),\\
& c(x,0)=q(x,0) = 0\, , && x\in(0,L),\\
& u c_\textrm{in} = \left(u c- D\pdv{c}{x}\right)_{x=0^+}\, ,\quad \left. \pdv{c}{x}\right|_{x=L^-}=0\,, \quad && t>0\,,
\end{alignat}
\end{subequations}
which, using the definition $D=d\mathcal{D}$, becomes the system \eqref{eq:1d}. We recall that $\mathcal{D}$ is the standard Brownian diffusion constant, while $d$ is computed from the integral in \eqref{eq:dij}. The same value of $d$ is obtained regardless of which of the three solutions ($i=1,2,3$) of the cell problem \eqref{eq:cell_c} is used. One may solve \eqref{eq:cell_c} numerically using commercial software or use the explicit formula given in \eqref{eq:d}.

The dimensional version of Darcy's law \eqref{eq:Darcy_sol}, can be obtained by reverting the change of variables \eqref{non_dim_Var} and the pressures scale \eqref{eq:p_nondim}, to give  
\begin{equation}\label{eq:Darcy_sol_dim}
    u  =-\frac{k l^2}{\phi\mu} \frac{(p_{\text{atm}}-p_0)}{L}=-\frac{\kappa}{\phi\mu} \frac{(p_{\text{atm}}-p_0)}{L}\,, 
\end{equation}
where the permeability of the porous media is $\kappa \equiv k l^2$, being $l$ the side-length of the cubic cell. The coefficient $k$ is computed by evaluating the integral \eqref{eq:u_int_2}, where the functions $K_{ij}$ are solutions of the cell problem given in~\eqref{eq:cell_up}. These problems can be readily solved using standard commercial software. Again, due to the symmetry of the formulation, any of the three diagonal components $K_{ii}$, with $i = 1, 2, 3$, may be used in the calculation, yielding the same result.

The side length $l$ can be expressed in terms of the porosity of the medium, $\phi$, and the diameter of the solid grains, $d_p$, leading to the following expression for the permeability of the medium 
\begin{equation}\label{Aguareles_Font}
\kappa =k\, d_p^2\,\left[\frac{\pi}{6(1-\phi)}\right]^{2/3}. \end{equation}
%$$\kappa =k r^2\left(\frac{4\pi}{1-\phi}\right)^{2/3}. $$
Note that the $d_p^2$ dependence of $\kappa$ appears in the widely used Kozeny–Carman expression, $\kappa = \phi^3d_p^2/[180(1-\phi)^2]$,
as well as in several other common permeability correlations for packed beds \cite{Perm_exp,Perm_review}. 

The expression in~\eqref{eq:Darcy_sol_dim} provides an approximation for the interstitial velocity in~\eqref{eq:1d_vis}. However, in practice, the inlet pressure $p_0$ is often unknown or not directly measured. In contrast, the inlet (or superficial) velocity of the flow, defined as $u_{\text{in}}=Q/A$, where $A$ is the inlet cross-sectional area, can be estimated since a volumetric flow rate $Q$ is typically imposed at the inlet of the column. In this case, the interstitial velocity can be approximated as 
\begin{equation}\label{u_from_superfcial}
u = \frac{u_{in}}{\phi}\,. 
\end{equation}
Expression \eqref{u_from_superfcial} is a popular choice for the interstitial velocity in 1D models like \eqref{eq:1d_vis} due to its simplicity and the fact that it does not require knowledge of permeability or inlet pressure, unlike Darcy's law \eqref{eq:Darcy_sol_dim}. 

In Section \ref{sec:assessment_1D}, we compare the predictions of the 1D model \eqref{eq:1d_vis} with computational results for microstructures that do not exhibit the same periodicity assumed in the homogenization process. In this case, since neither the inlet pressure nor the permeability is known, we use the interstitial velocity \eqref{u_from_superfcial}. We demonstrate that, even when the periodicity of the computational microstructure differs from that used in the homogenization, and even when the spheres forming the microstructure are of moderate size (comparable to the cylinder radius), the 1D model provides an exceptionally accurate description of the process. In Section \ref{sec:compu_exp_2}, we design computational experiments in which the microstructure exactly matches that used in the homogenization process. In this case, both the permeability and inlet pressure are known, so the interstitial velocity is computed using \eqref{eq:Darcy_sol_dim}. We then demonstrate how the computational solution converges to the solution of the 1D model \eqref{eq:1d_vis} as the microstructure is refined, increasing the number of cells but keeping the porosity constant.

%Note that the homogenized model and its corresponding 1D reduction were formulated for a microstructure consistent in a periodic array of spheres, where the size of the unit cell is much smaller than the adsorption length-scale. This structure might look rather restrictive and far from the actual randomly distributed grains that one finds in real applications. However, in the following sections we will show how, even when the arrays are not periodic arrays of spheres in cubic cells, and/or the spheres  forming the microstructure have moderate sizes comparable with the cylinder radius, the solution of the derived 1D model gives an extremely accurate description of the process, which proves the robustness of the 1D reduction. 

\section{3D computational experiments and assessment of the averaged 1D model}
\label{sec:asses}

In this section, we develop computational experiments to: (i) analise the effect of the microstructure of the porous media in the adsorption process and (ii) asses the capacity of the averaged 1D model to describe the results of the 3D computational experiments. The experiments involve the solution of the 3D model developed in Section~\ref{sec:model} on idealized porous structure domains, using operating conditions from real column adsorption experiments extracted from the literature. The results of the computational experiments are then compared to the predictions of the 1D model derived in Section~\ref{sec:1D_reduction}.  

\subsection{Computational experiments: parameters and geometries}

To build the computational experiments, we take as a reference the column adsorption experiments for CO$_2$ capture in He-CO$_2$ mixtures developed in \cite{Delgado2006}. We focus on the experimental runs performed at low volume fraction of CO$_2$ in He, in particular the ones where only the 14\% of the mixture volume is composed of CO$_2$, such that the diluted approximation is not challenged. The column geometry and parameter values that enter our 3D model \eqref{eq:model_ad}-\eqref{eq:model_fluid} are listed in Table~\ref{tab:Delgado}. 

\begin{table}[H]
    \centering
    \begin{tabular}{|l|c|c|c|}\hline
        Property & Variable & Units & Value \\
        \hline
        Inlet CO$_2$ concentration & $c_{in}$ & mol/m$^3$ & 5.436  \\
        Max. adsorbed fraction & $q_{max}$ & mol/kg & 4.975  \\
        Inlet velocity & $u_{\text{in}}$ & m/s & 1.69$\times10^{-3}$ \\
        Column length & $L$ & cm & $16.3$ \\
        Column radius & $R$ & cm & $0.8$ \\
        Void fraction & $\phi$ & - & $0.52$ \\
        Adsorbent particle density & $\rho_a$ & kg/m$^3$ & $1070$ \\
        %Eq. constant & $K$ & m$^3$/mol & $0.01096$ \\
        Adsorption constant & $k_{ad}$ & m$^3$/mol$\cdot$s & 4.72$\times10^{-4}$\\
        Desorption constant & $k_{de}$ & s$^{-1}$  & 4.31$\times10^{-2}$ \\
        Partial orders & $m,\,n$ & - & 1,\,1 \\          
        Gas molecular diffusivity & ${\cal D}$ & m$^2$/s & 0.65$\times10^{-4}$\\
        Gas viscosity & $\mu$ & Pa$\cdot$s & 1.8$\times10^{-5}$ \\  
        %Gas mix. density  & $\rho$ & kg/m$^3$ & 0.36 \\ 
        \hline
    \end{tabular}
    \caption{Material properties and operating conditions used in the computational experiments. The first seven properties are extracted from the experiments reported in \cite{Delgado2006}. The adsorption/desorption constants and the partial orders of the sorption process are extracted from \cite{Myers24}. The last two are from the database \cite{DataBaseColorado}. }
    \label{tab:Delgado}
\end{table}

\begin{figure}[htb]
\begin{center}
\includegraphics[width=0.5\textwidth]{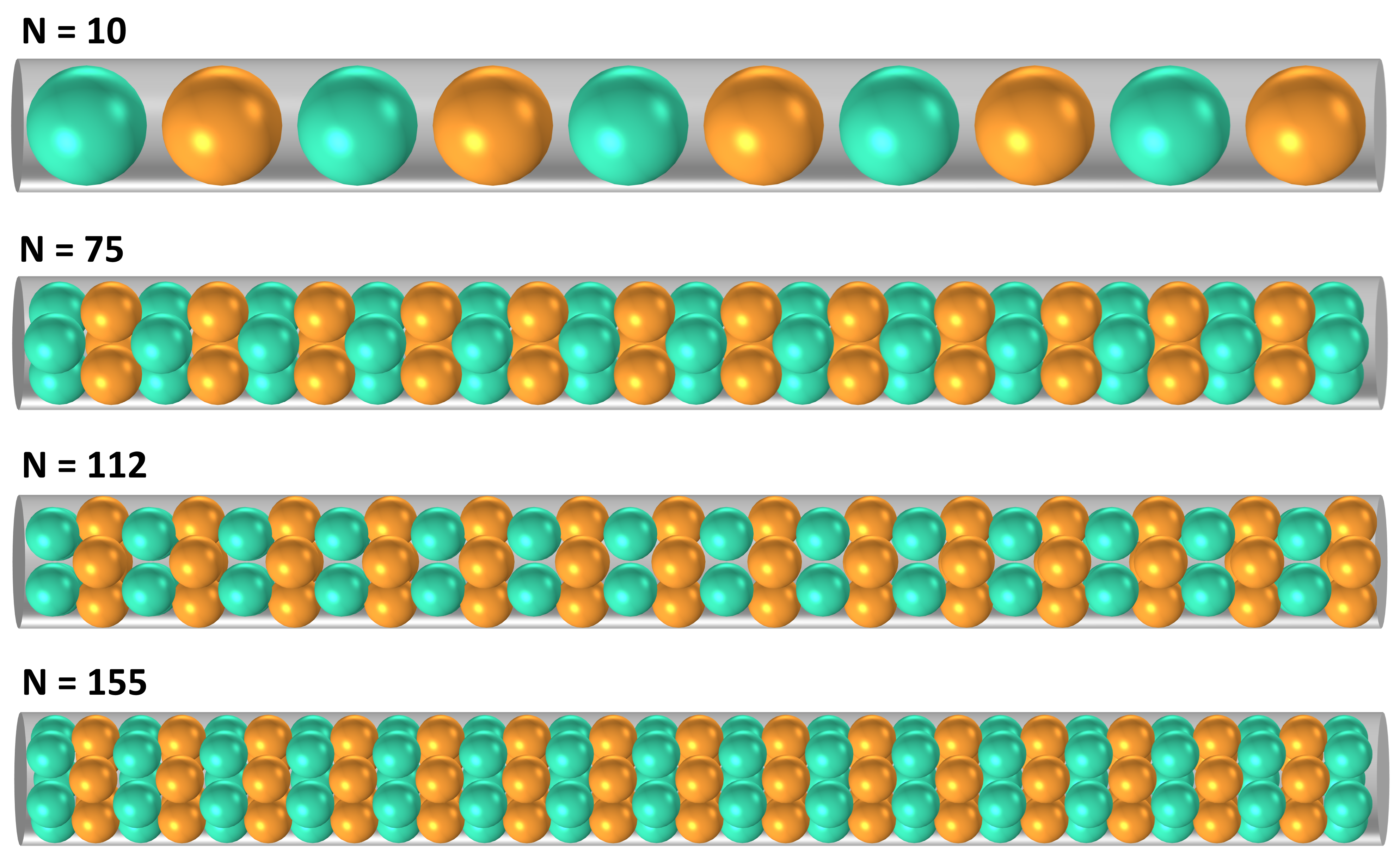}
\end{center}
\caption{Lateral view of the geometries used in the computational experiments. The inlet of the column is located on the left end, $z=0$. The different colors indicate different layers. The packing of spheres is formed by a unit cell, a green and a yellow layer, that repeats throughout the column. }
\label{fig:cases1}
\end{figure}

\begin{figure}[htb]
\begin{center}
\includegraphics[width=0.5\textwidth]{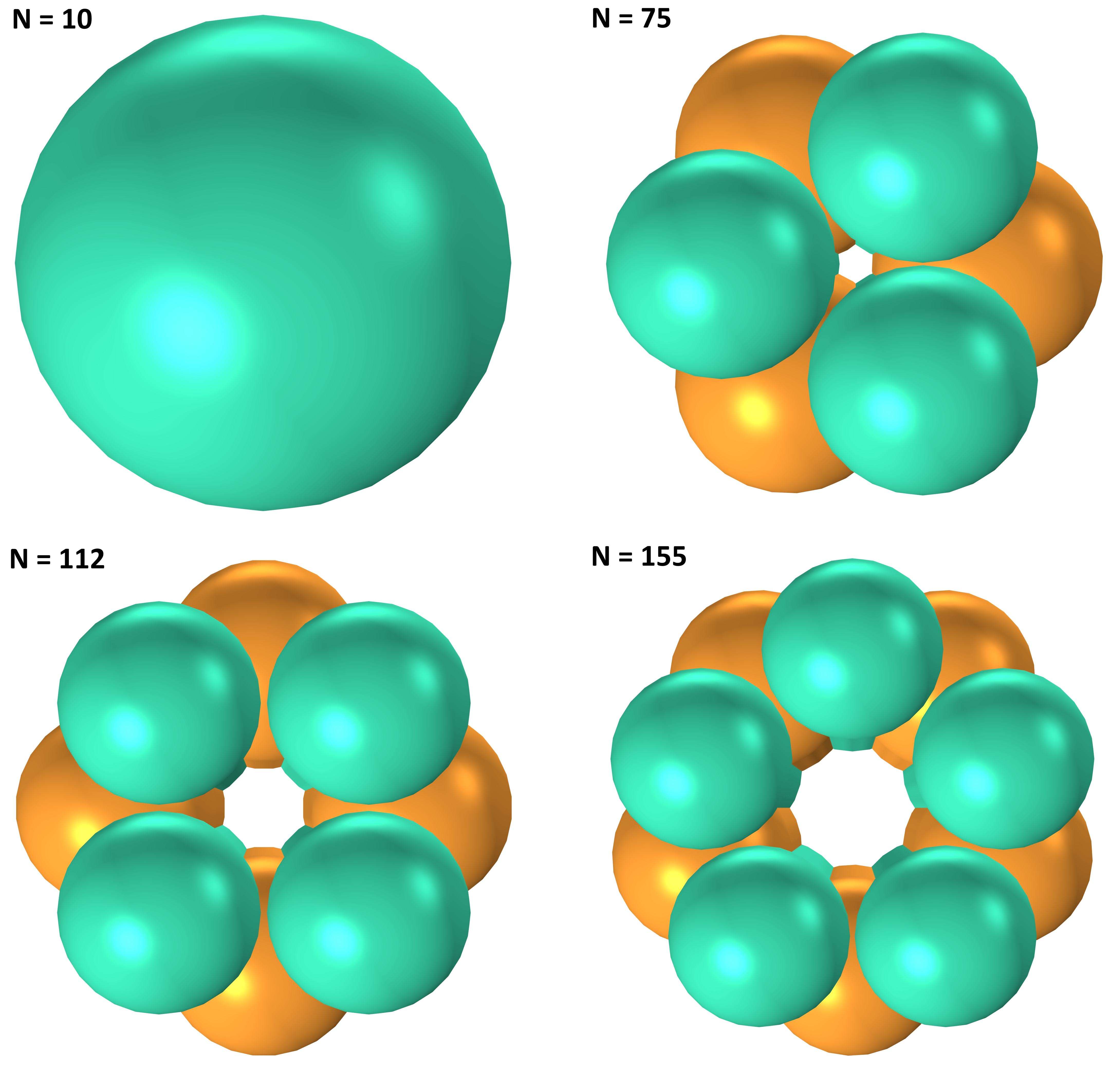}
\end{center}
\caption{Crossectional view of the microstructures used in the computational experiments. }
\label{fig:cases2}
\end{figure}

The geometries and the numerical solutions of the 3D model are performed with the commercial software COMSOL Multiphysics. The macroscopic geometry of the computational experiments consists of a cylindrical column of radius $R$ and length $L$ (see values in Table~\ref{tab:Delgado}). To simulate the microstructure of the porous media, the interior of the column is filled with a dense arrangement of spherical particles following the approach described in \cite{Mughal12}. This involves generating a 3D optimal packing of spheres by first looking at the 2D optimal packing of $n$ circles in a larger circle of radius $R$. By doing this, we can build up the 3D arrangement layer by layer, ultimately achieving a dense 3D packing of spheres within the cylinder. We consider 4 diferent cases, using $n=1, 3, 4, 5$ spheres per layer, as shown in Figures~\ref{fig:cases1} and \ref{fig:cases2}, which make for a total of $N=10, 75, 112, 155$ particles within the cylinder, respectively. Once the arrangement is obtained, the radius of the particles is slightly decreased to match the experimental porosity $\phi=0.52$ in all cases. This also avoids contact between contiguous spheres, thereby preventing potential numerical issues at the contact points. The final particle radius ($r_p$) for each case and the corresponding angle of rotation ($\theta$) between one layer and the next are summarized in Table~\ref{tab:micro1}. Note the case $n=2$ has not been included since a porosity of $\phi = 0.52$ cannot be achieved for the dimensions $R$ and $L$ of the column used in the present study.  

\begin{table}[htb]
\centering
    \begin{tabular}{|c|c|c|c|}
    \hline
    $N$ & $n$ & $\theta$ & $r_p$ [m] \\ 
    \hline 
    10  & 1 &  -      & 0.007195 \\
    %\hline
    75  & 3 & $\pi/3$ & 0.0036757 \\
    %\hline 
    112 & 4 & $\pi/4$ & 0.0032176 \\
    %\hline 
    155 & 5 & $\pi/5$ & 0.0028875 \\
    \hline 
    \end{tabular}
    \caption{Parameter values for the 4 microstructures  studied in this work, where $N$ is the total number of particles, $n$ is the number of particles in each layer, $\theta$ is the rotation angle between layers, and $r_p$ is the particles's radius.}
    \label{tab:micro1}
\end{table} 

The gas mixture flows into the column through the inlet positioned at its left end (see Fig.~\ref{fig:cases1}) at a speed $u_\text{in}$ and with a contaminant concentration of $c_{\text{in}}$ (see Table~\ref{tab:Delgado}). The gas is allowed to leave the column through the outlet positioned at its right end. While flowing through the free interparticle space inside the column the contaminant is progressively adsorbed on the surface of the particles. When the adsorbed fraction on the particles surface reaches its equilibrium capacity everywhere, the adsorption process ends. At this point, the outlet concentration of contaminant will match the inlet value. In the next section, we describe and discuss the flow velocities and the evolution of the concentration profiles obtained in the four experiments.  

%In the next section, we specify the solution procedure followed using COMSOL    

\subsection{Results of the computational experiments} 

To solve the 3D model \eqref{eq:model_ad}-\eqref{eq:model_fluid} numerically in COMSOL, we use the \textit{Transport of Diluted Species} option for \eqref{eq:MB} and the \textit{Creeping Flow} option for \eqref{eq:stokes}. These equations are solved in $\Omega_f$, i.e. the free space between the spherical particles and the column. Since no pre-established option exists to model adsorption on surfaces, we introduce equation \eqref{eq:adsorption_vis} via the Mathematics module, in particular we use the option \textit{General Form Boundary PDE} to define and solve \eqref{eq:adsorption_vis} on the particles surface, $\partial\Omega_s$. To ensure a high accuracy in the solutions, the meshes are automatically generated with the \textit{Physics-controlled-mesh} option, choosing a \textit{Fine} element size, and a relative tolerance for the time-dependent solver of 10$^{-5}$. All simulations were run on a workstation equipped with a 13th Gen Intel Core i9-13900K CPU at 3.00 GHz and 128 GB of RAM. Note that in  \eqref{eq:model_ad}-\eqref{eq:model_fluid} the velocity equation is decoupled from the concentration and adsorption equations. This allows to solve for the velocity field first and use the solution as input value in the concentration equation, which leads to relatively fast computing times. In particular, the longest runtime occurred for the case $N=155$, with a computation time of 6 minutes for the velocity field and 38 minutes for the concentration.  

\begin{figure}[htb]
\begin{center}
\includegraphics[width=0.5\textwidth]{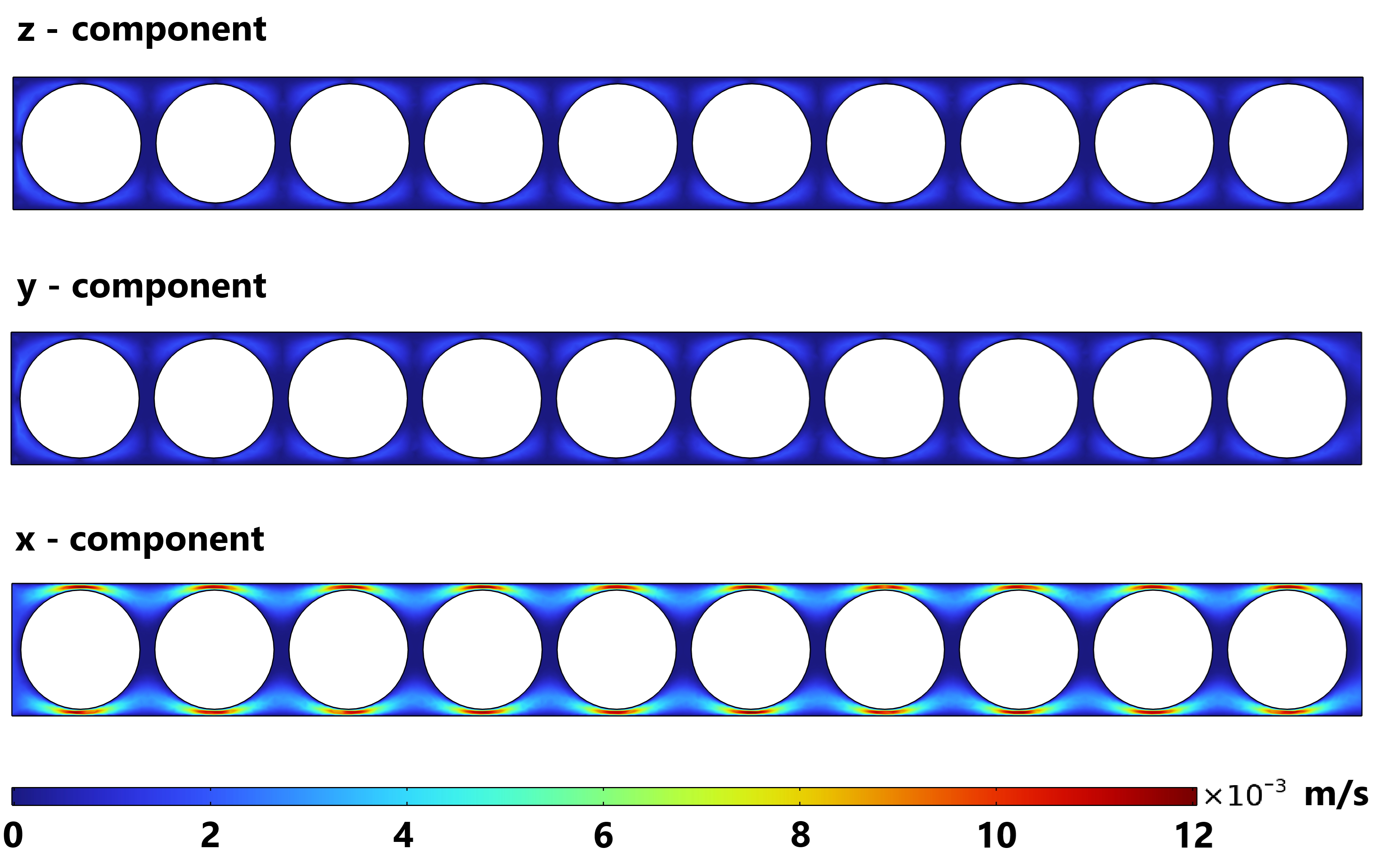}
\end{center}
\caption{Velocity field along a longitudinal plane for the case $N = 10$. }
\label{fig:velocity_2D}
\end{figure}

In Figure~\ref{fig:velocity_2D} we present the velocity field for the case $N=10$ along a longitudinal plane. The magnitudes of the $z$ and $y$ components of the velocity field are very small when compared to the magnitudes of the $x$ component. Focusing on the $x$ component, we observe that the velocity is much larger on the regions where the spheres are closer to the column wall. This behavior has two origins. First, it arises in regions with smaller cross-sections, which necessitate faster flow to maintain mass conservation under the incompressibility condition \cite{Incropera}. %Second, the flow pattern around each sphere exhibits the characteristic behavior of flow over a sphere \cite{Incropera}: low velocities at the front and rear, and high velocities along the sides. 

\begin{figure}[htb]
\begin{center}
\includegraphics[width=0.5\textwidth]{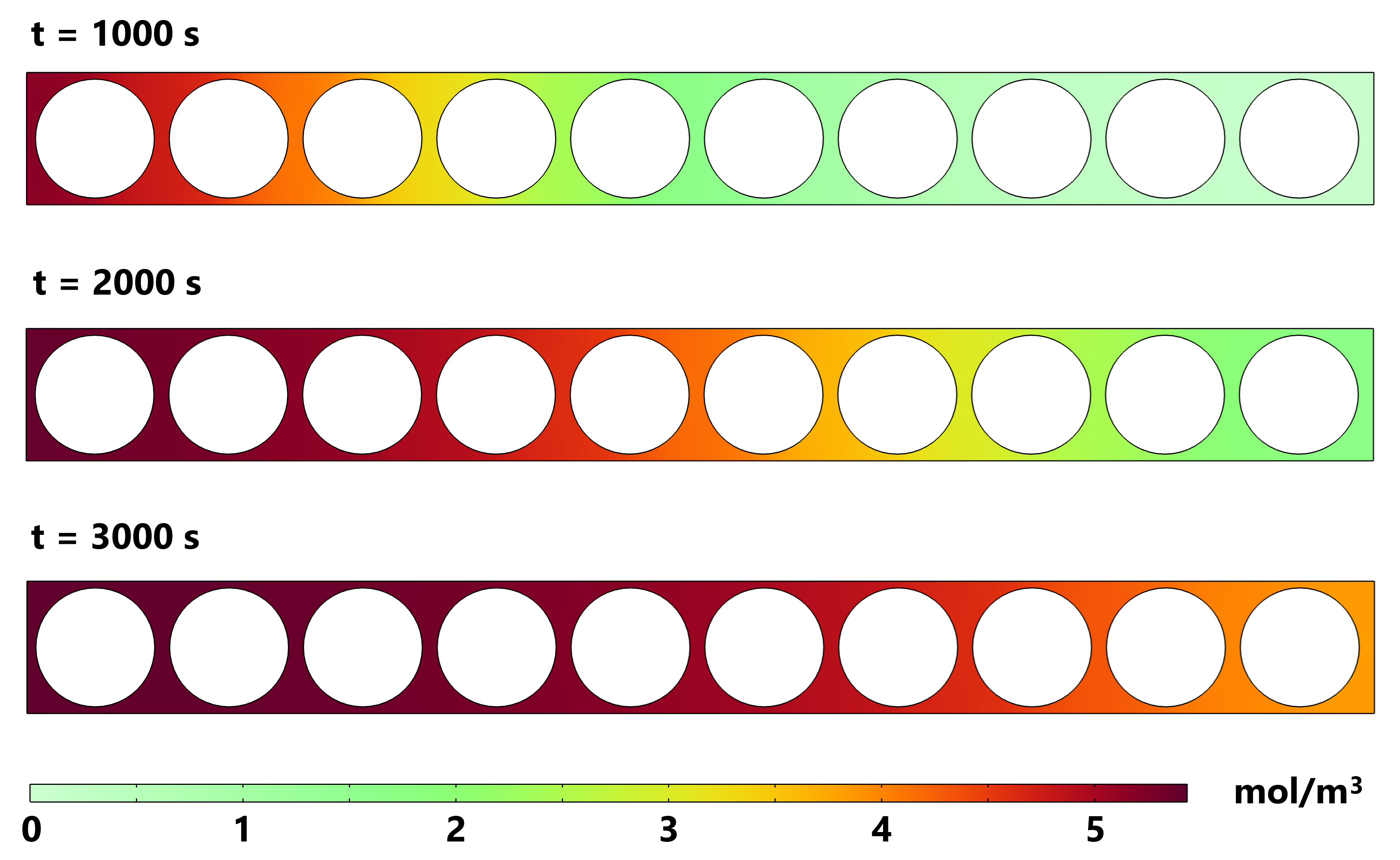}
\end{center}
\caption{Concentration color maps for the case $N = 10$ at times $t =$  1000 s, 2000 s, 3000 s.  }
\label{fig:concentration_2D}
\end{figure}

In Figure~\ref{fig:concentration_2D} we present the evolution of the concentration of CO$_2$ within the column for the case $N=10$. The progression of CO$_2$ reassembles that of a planar front wave propagating from left to right. This traveling wave behavior, is a characteristic feature of column adsorption problems which has been extensively studied in 1D models by the authors and collaborators (see, for instance, \cite{Myers20a,SipsPaper,Myers24,angladalloveras}). 

\begin{figure}[htb]
\begin{center}
\includegraphics[width=0.5\textwidth]{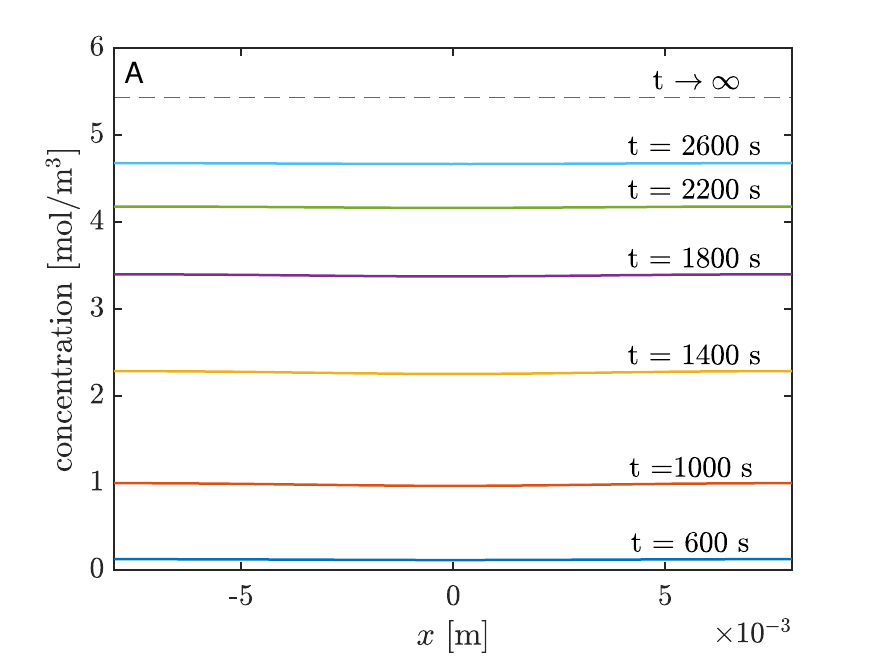}\includegraphics[width=0.5\textwidth]{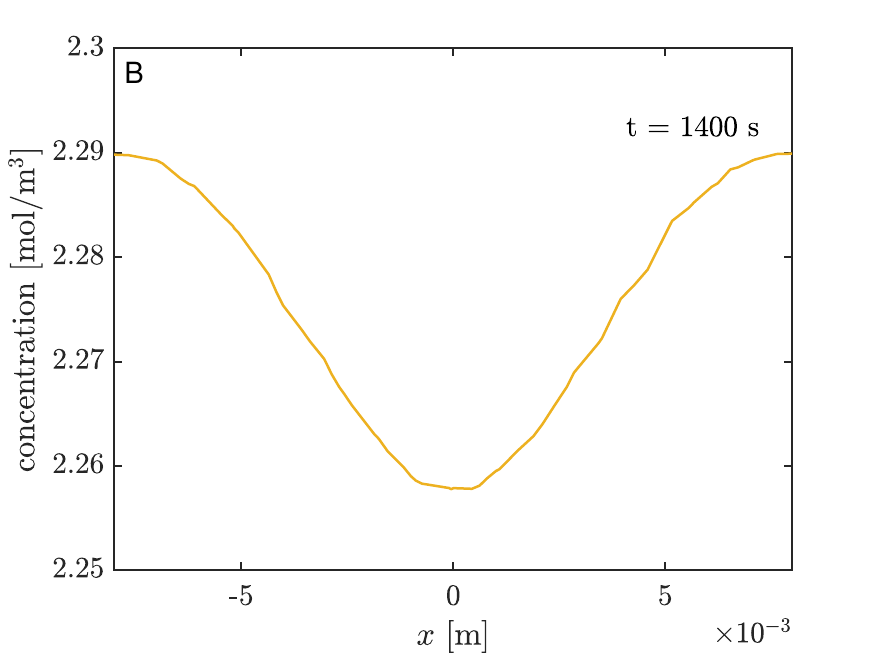}
\end{center}
\caption{Panel A shows the concentration profiles at the centre of the column for the case $N = 10$ at different times. These correspond to profiles along the line of length $2R$ that goes from $(x,y,z) = (L/2,-R,0)$ to $(x,y,z) = (L/2,R,0)$. The dashed line indicates the inlet concentration. Panel B is a zoom of the concentration profile at $t=1400$ s.}
\label{fig:profiles_crossection}
\end{figure}

Since any 1D approximation to the problem involves some degree of averaging over a cross section of the column, an important aspect to analyse from the 3D computational experiments is the radial distribution of contaminant. By looking at Figure~\ref{fig:concentration_2D}, we observe that the primary variation in contaminant distribution occurs along the axial direction, while the radial direction exhibits a uniform contaminant distribution. To further analyse the radial distribution of contaminant, in Figure~\ref{fig:profiles_crossection} we show the evolution of the concentration on a line perpendicular to the $x$ axis that crosses the column between particles number 5 and 6 (i.e., at the center of the column). Panel A clearly shows that the evolution of the concentration is an almost flat profile that increases in time. We note that the configuration for $N=10$ is radially symmetric, so the evolution on this line is representative of the behaviour on the cross section. A zoom in of the profile at $t=1400$ s is shown on panel B, indicating a tiny difference of approximately 2.29-2.26 = 0.03 mol/m$^3$ in the concentration between the center and the wall of the column. The minimal variation in concentration across the column's cross sections observed in the 3D experiments validates the use of averaged 1D models under the assumptions of the current study specifically, laminar incompressible Stokes flow and the dilute species approximation. 

The general features of flow velocity and concentration observed for the case $N=10$ persist in the cases $N=75, 112, 155$, so we skip the discussion of these cases. The equivalent velocity and concentration plots for $N=75, 112, 155$ are shown in the Supplementary Information. 

In the next section, we analyse the differences between the prediction of the 1D and 3D models, focusing on the concentration profiles within the column and the evolution of the concentration at the column outlet. 

\subsection{Assessment of the one-dimensional model}
\label{sec:assessment_1D}

One of the most important measurements used in column adsorption experiments is the concentration of contaminant at the outlet of the column \cite{Gabelman2017}. The time evolution of the concentration at the outlet is named the breakthrough curve. Fitting the solution of a 1D theoretical model to the experimental breakthrough curve is the main procedure by which the adsorption/desorption constants for a particular column experiment are found (see, e.g., \cite{SipsPaper}, \cite{Myers24}). Therefore, testing the limits and assessing the validity of 1D models is critical for accurately predicting the controlling parameters in column adsorption experiments.  

In this section, we compare breakthrough curves and concentration profiles obtained from the 1D model and the 3D computational experiments. It is important to note that the microstructure assumed in the homogenization procedure differs significantly from that used in the computational experiments. While both approaches model the adsorbent material using spherical particles, the homogenization framework assumes a periodic array of spheres in cubic cells, with diameters much smaller than the adsorption length. In contrast, the computational experiments employ spheres that do not follow the same periodicity and are considerably larger. Rather than attempting to compare two approaches designed to solve the exact same problem, our goal in this section is to demonstrate that, even when the porous medium consists of relatively coarse structures, the 1D model, based on a fine periodic structure, can still provide remarkably accurate predictions of the process.

The dimensionless parameters of the 1D model \eqref{eq:1D-B} are found using the parameters and experimental conditions in Table~\ref{tab:micro}: $\text{Da} = 1.1\cdot10^{-3}$, $\text{Pe}^{-1} = 14.34$ and $\delta = 16.78$. Solving the cell problem \eqref{eq:cell_c} numerically with COMSOL and computing the integrals involved in \eqref{eq:A} we obtain $\text{Pe}_{\text{eff}}^{-1} = 11$. Since the inlet pressure is unknown and the microstructure differs from that used in the homogenization, we have used equation \eqref{u_from_superfcial} instead of \eqref{eq:Darcy_sol_dim} to compute the flow velocity in the 1D model. The solution of the 1D model is obtained numerically implementing a standard forward time and centered in space finite difference scheme in a home-made Matlab code. The authors have used this implementation, or slight variations of it, in several previous works on contaminant capture (see \cite{Myers20a,Myers20b,Myers24}, to name a few). Given that the numerical scheme is well-established and poses no significant challenges, we omit a detailed description.

\begin{figure}[htb]
\begin{center}
\includegraphics[width=0.5\textwidth]{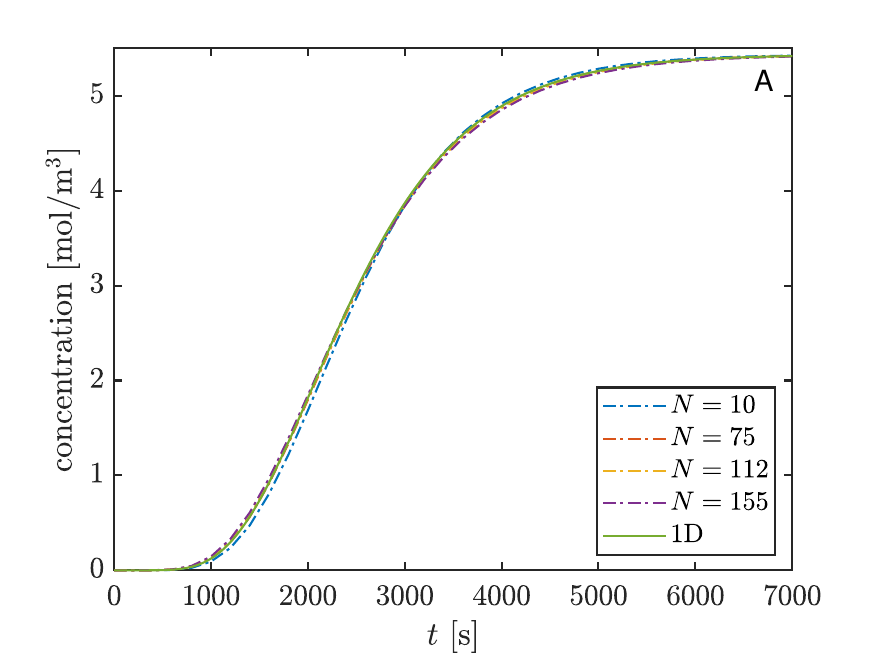}\includegraphics[width=0.5\textwidth]{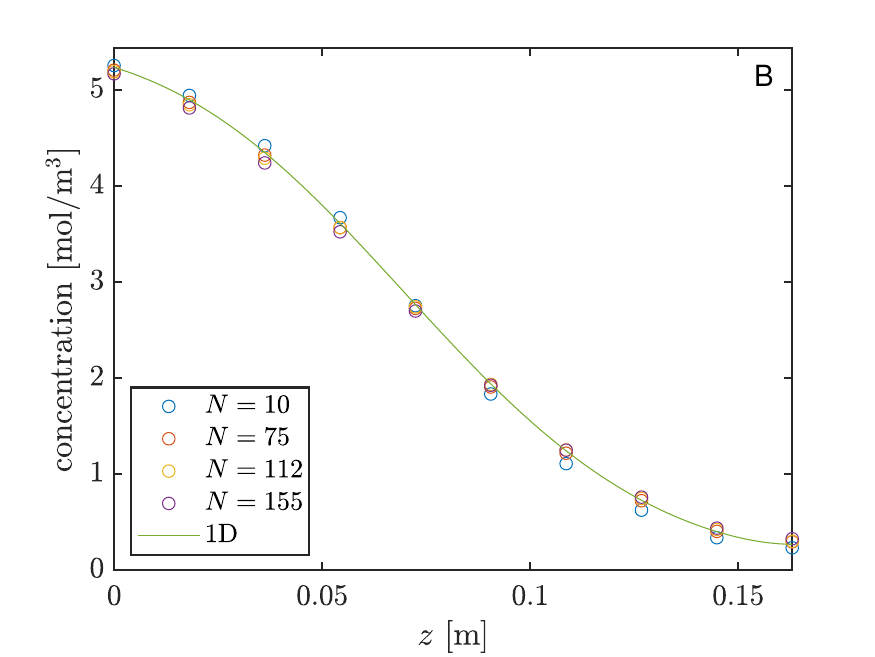}
\end{center}
\caption{Panel A shows the breaktrhough curves for the 3D simulations and the 1D model. The breakthrough curves for the simulations are obtained by averaging over the outlet crossection. Panel B shows the concentration profiles along the $z$ axis at $t = 1400$ s for the 3D simulations and the 1D model. The profiles from the 3D simulations correspond to column crossection averages at 10 equispaced  points in the $x$ axis, i.e. $x= p\,L/9$ for $p=0,\ldots,9$. }
\label{fig:profiles_crossection_BT}
\end{figure}

\begin{figure}[htb]
\begin{center}
\includegraphics[width=0.5\textwidth]{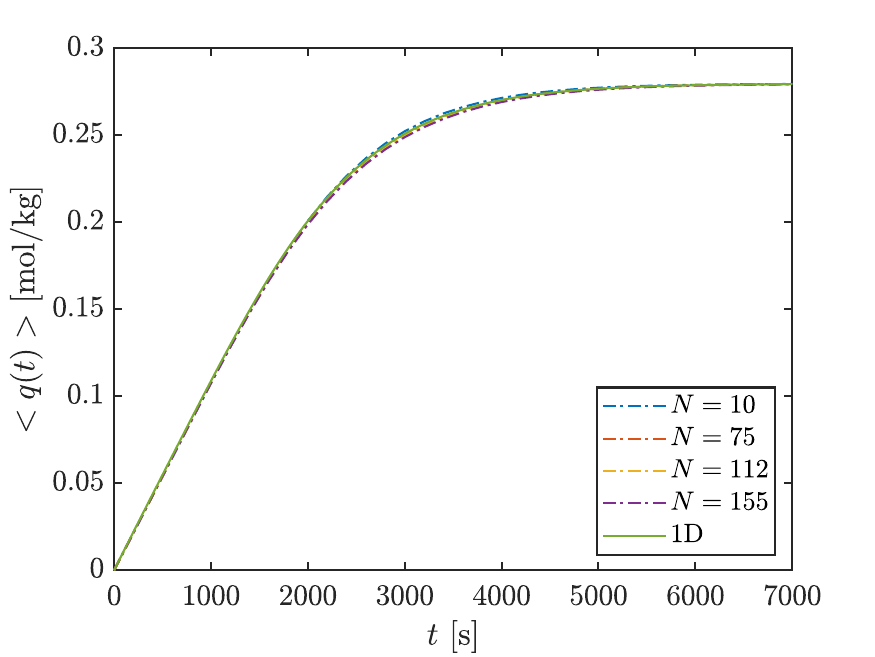}
\end{center}
\caption{Evolution of the averaged adsorbed fraction over the particle's surface for each case ($N=$ 10, 75, 112, 155) and comparison to the averaged adsorbed fraction from the 1D model. }
\label{fig:adsorbed_fraction}
\end{figure}

Panel A of Figure~\ref{fig:profiles_crossection_BT} presents the breakthrough curves obtained from the 3D computational experiments conducted in this study along with the breakthrough curve from the 1D model. The breakthrough curves from the computational experiments are obtained by averaging the concentration over the outlet cross section. The breakthrough curves for all experiments exhibit remarkable similarity, indicating that the geometry's microstructure has minimal impact on the overall adsorption process. Moreover, the breakthrough prediction from the 1D model closely matches those from the computational experiments, with the case $N=10$ showing a slightly greater divergence than the others. 

Panel B of Figure~\ref{fig:profiles_crossection_BT} compares the concentration profile predicted by the 1D model with the results of the computational experiments along the column at $t=$ 1400 s. The circles correspond to cross section averages of the concentration for the four computational cases studied at 10 equispaced $yz$-planes within the column. The solid line represents the prediction of the 1D model. Once again, the 1D solution effectively captures the trend of the computational experiments, showing a slightly less precise match with the case $N=10$ compared to the rest.

We now analyse the evolution of the amount of contaminant adsorbed within the column throughout the process. In Figure~\ref{fig:adsorbed_fraction}, we show the evolution of the average of the adsorbed fraction over the particle's surface, $<q(t)> = |\partial\Omega_s|^{-1}\int_{\partial  \Omega} q(t,\boldsymbol{x})\, dS$. In the case of the 1D model, the average reduces to $<q(t)> = L^{-1}\int_{0}^{L} q(t,x)\, dx$. Similar to the concentration profiles and the breakthrough curves, the differences between the four computational experiments and the 1D model are tiny, strongly suggesting that the microstructure of the media has a tiny impact on the overall adsorption process. 

\subsection{Analysis of the convergence to the homogenised model}
\label{sec:compu_exp_2}

The low porosities in the reference experiments used in the previous sections \cite{Delgado2006} did not permit the use of the same microstructure as that assumed in the homogenised model. The low porosity ($\phi = 0.52$) necessitated a denser packing of spheres than that offered by a simple periodic array of spheres centered in cubic unit cells. While the 1D solution successfully captured the overall trends observed in the 3D simulations, the comparison remained somewhat indirect due to the mismatch between the assumed microstructures.

\begin{figure}[htb]
\begin{center}
\includegraphics[width=1\textwidth]{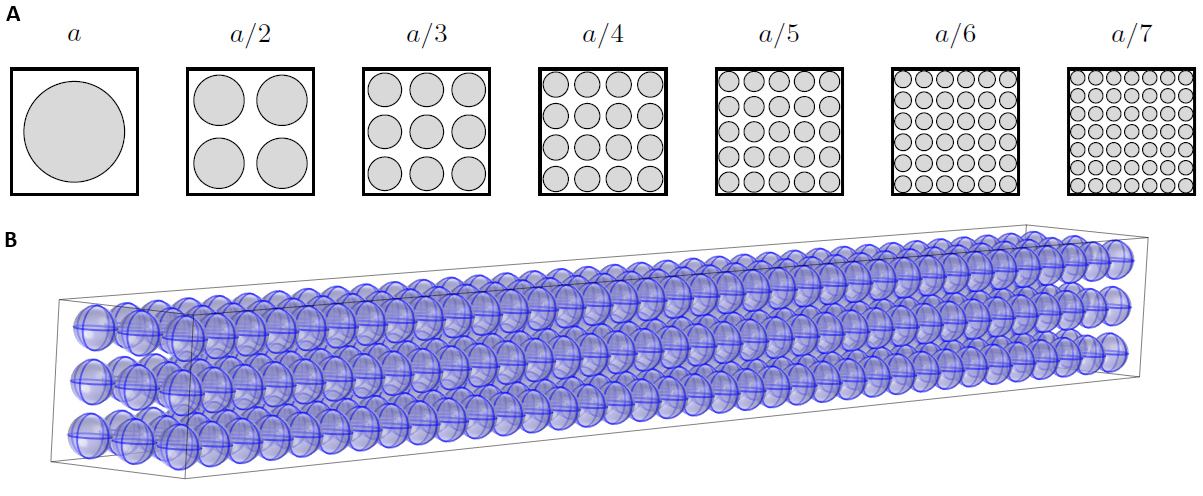}
\end{center}
\caption{The sequence A illustrates the front view of the different computational cases studied in Section~\ref{sec:compu_exp_2} as cell size $l$ decreases and B shows an illustration of the whole square column for the case $l=a/3$.}
\label{fig:square_column}
\end{figure}

To address this issue, in this section we perform a new set of 3D computational experiments using a geometry that exactly replicates the periodic array of spheres employed in the homogenisation procedure. Specifically, we consider a column with a square cross-section (side length $a = 0.014$ m) and total length $L=12a$, which allows a perfect tiling of the unit cells. The porosity is set to $\phi = 0.7$ to accommodate this idealised packing and $u_{in} = 2.26\times10^{-3}$ m/s, while all other model parameters are kept identical to those in the previous computational experiments (see Table~\ref{tab:Delgado}). We consider seven configurations, each corresponding to a periodic array of spheres within the square column, where the side length of the repeating unit cell, $l$, is progressively refined. The first case uses a single unit cell of side length $l=a$; the second uses a cell of size $l=a/2$; the third, $l=a/3$; and so on, down to $l=a/7$. The side length, $l$, the particle diamenter, $d_p$, and the total number of particles in the column, $N$,  for each case are listed in Table~\ref{tab:micro}. An illustration of the front view for the different cases and, as a representative example, the geometry for the case with unit cell size $l=a/3$ are shown in Figure~\ref{fig:square_column}. These configurations enable a direct and unambiguous comparison between the 1D and 3D models, and allow us to examine the convergence of the 3D results toward the 1D homogenised solution as $N$ increases.

\begin{table}[htb]
\centering
    \begin{tabular}{|c|c|c|c|c|c|c|}
    \hline
    $N$ & $l$ ($a = 0.014$ m) & $d_p$ [m] & $\kappa$ ($\times10^{-5}$) [m$^2$] & $p_0-p_\text{atm} [Pa]$ & $u$ [m/s] & $\Delta u$ [m/s] \\ 
    \hline       
    12   & $a$     & 0.011628 & 0.2083 & 0.017845 & 0.0176 & 0.0142  \\ 
    96   & $a/2$   & 0.005814 & 0.0521 & 0.025397 & 0.0063 & 0.0029 \\
    324  & $a/3$   & 0.003876 & 0.0231 & 0.042094 & 0.0046 & 0.0012 \\
    768  & $a/4$   & 0.002907 & 0.0130 & 0.064746 & 0.0040 & 0.0006 \\
    1500 & $a/5$   & 0.002326 & 0.0083 & 0.094224 & 0.0037 & 0.0003 \\
    2592 & $a/6$   & 0.001938 & 0.0058 & 0.12949  & 0.0035 & 0.0001 \\
    \textbf{4116} & $\boldsymbol{a/7}$ & \textbf{0.001661} & \textbf{0.0043} & \textbf{0.17083}  & \textbf{0.0034} & - \\
    \hline 
    \end{tabular}
    \caption{Parameter values for the 7 microstructures studied in Section~\ref{sec:compu_exp_2}. The first column is the number of inclusions $N$. The last column indicates the absolute error computed as $\Delta u = |u_{N=4116} - u_{N}|$, where $N \in \{12, 96, 324, 768, 1500, 2592\}$.}
    \label{tab:micro}
\end{table}

\begin{figure}[htb]
\begin{center}
\includegraphics[width=0.5\textwidth]{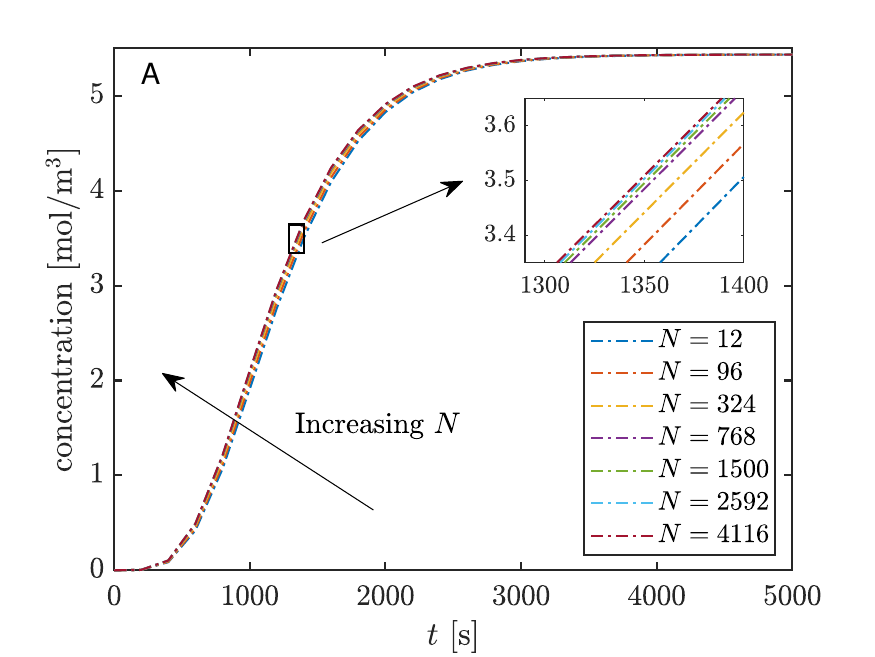}\includegraphics[width=0.5\textwidth]{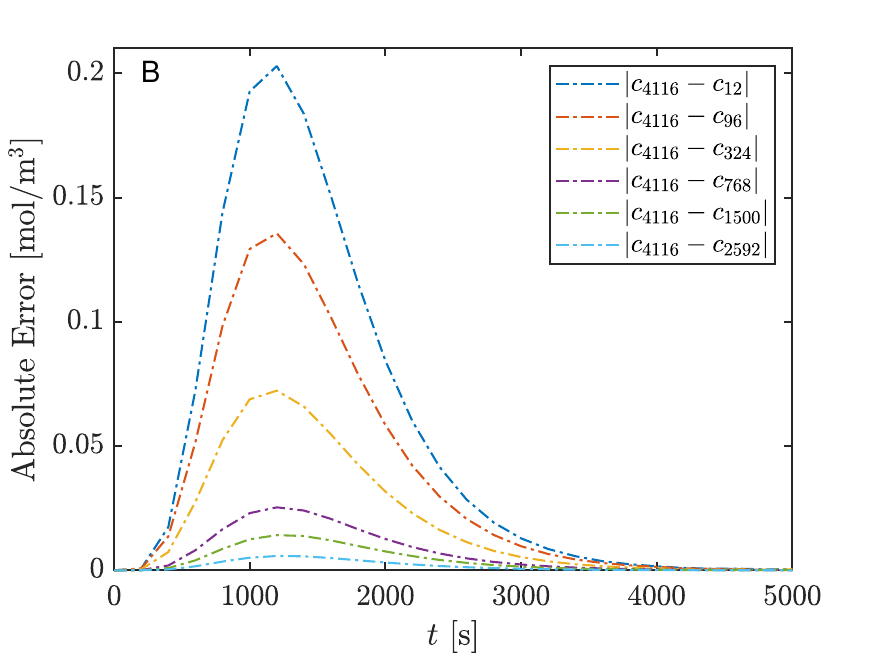}
\end{center}
\caption{Panel A shows the breakthrough curves for the computational cases $N = 12, 96, 324, 768, 1500, 2592, 4116$ (i.e., cell sizes $l = a,a/2,a/3,a/4,a/5,/a/6,a/7$). The inset displays a zoomed-in view of the breakthrough curves, highlighting the leftward shift as $N$ increases. Panel B shows the evolution of the absolute error of the breakthrough curves in Panel A taking as reference the case $N = 4116$. }
\label{fig:square_increase_N}
\end{figure}

Panel A of Figure~\ref{fig:square_increase_N} presents the breakthrough curves from all computational cases examined in this section. Although the differences between the curves are subtle, we observe a slight leftward shift in the slope of the breakthrough curve as the number of particles, $N$, increases. The inset in Panel A further highlights that this shift becomes progressively smaller with increasing $N$, indicating a diminishing difference between the breakthrough curves. To quantify the differences between the breakthrough curves, we compute the absolute error of each curve relative to the one obtained with the largest number of particles, 
$N$. The steadily decreasing error clearly indicates that the breakthrough curves converge toward a limiting behavior as 
$N$ increases.

\begin{figure}[htb]
\begin{center}
\includegraphics[width=0.5\textwidth]{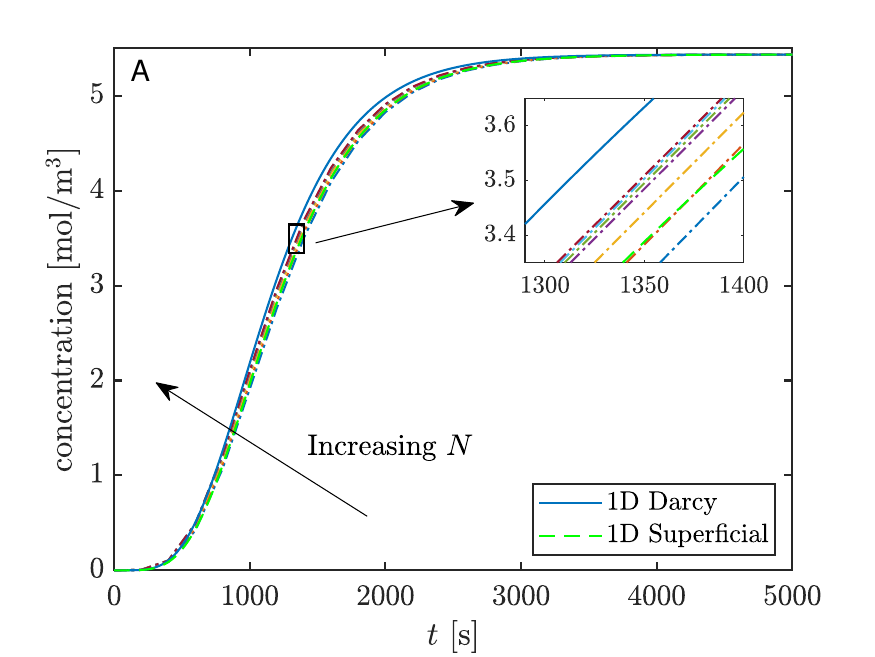}\includegraphics[width=0.5\textwidth]{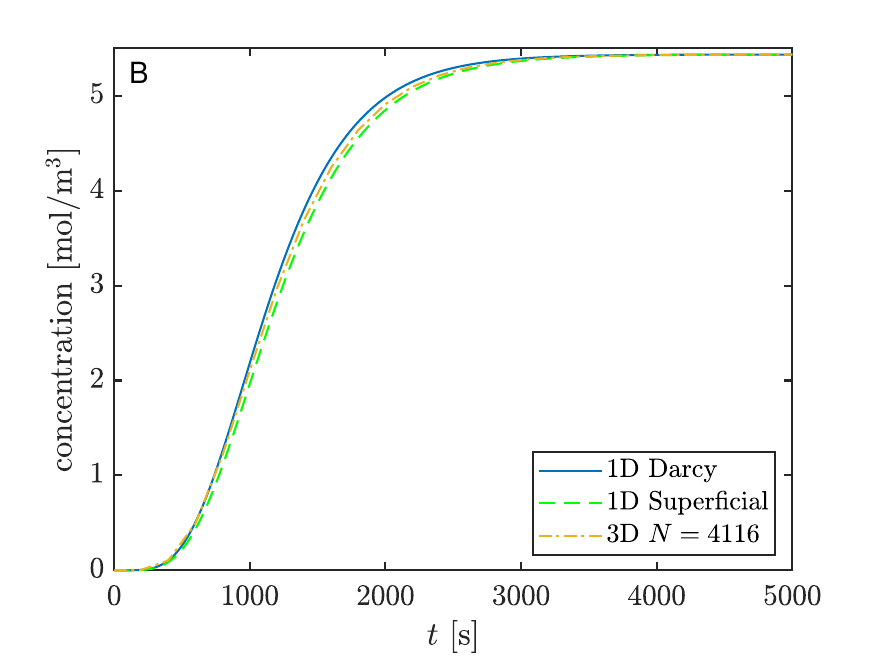}
\end{center}
\caption{In Panel A we overlap the solution of the 1D models, using Darcy velocity (solid line) and superficial velocity (dashed line) approaches, to the computational breakthrough curves shown in Figure~\ref{fig:square_increase_N}. In Panel B we show the evolution of the breakthrough curve for the case with the largest number of particles ($N=4116$) and the solution of the 1D models taking the two different approaches for the fluid velocity.} 
\label{fig:last}
\end{figure}

We now examine whether the computational experiments converge to the homogenised 1D model \eqref{eq:1d_vis}-\eqref{eq:Darcy_sol_dim} as the number of particles $N$ increases. The first step is to compute the flow velocity. Since the velocity expression \eqref{eq:Darcy_sol_dim} depends on the permeability, which itself depends on the particle radius, and since the inlet pressure $p_0$ varies across simulations, the resulting velocity $u$ differs for each case.

To compute the permeability, $\kappa$, whose expression is given in equation \eqref{Aguareles_Font}, we first determine the homogenisation coefficient $k$, which is obtained by solving the cell problem \eqref{eq:u_int_2} with a fixed porosity $\phi = 0.7$. This problem is solved numerically in COMSOL, yielding $k = 0.010628$. The permeability for each simulation is then computed based on the corresponding particle diameter $d_p$. As the average pressure at the inlet is directly accessible from the simulations, the pressure drop $p_{0} - p_{\text{atm}}$ can be readily calculated for each case. With both the permeability and pressure drop known, the interstitial velocity \eqref{eq:Darcy_sol_dim} can then be computed accordingly. These values are presented in Table~\ref{tab:micro}. Notably, the computed values of $u$ show clear signs of convergence toward a limiting value as $N$ increases, as indicated by the decreasing error $\Delta u$ (see last column of Table~\ref{tab:micro}).

Panel A of Figure~\ref{fig:last} compares the breakthrough curves from the 1D model, using either $u=u_{in}/\phi = 0.0032$ m/s (dashed line) or $u=0.0034$ m/s corresponding to the Darcy velocity for the case $N=4116$ in Table~\ref{tab:micro} (solid line), with the computational results for all 
$N$. The 1D solution using Darcy velocity lies slightly to the left of the $N=4116$ curve, consistent with the trend that increasing $N$ shifts the computational curves leftward. This suggests that the computational results approach the Darcy-based 1D model as  $N$ increases. Notably, the curves do not converge toward the 1D solution using superficial velocity, in fact, they diverge from it with increasing $N$ (see inset). For clarity, Panel B isolates the $N=4116$ case alongside both 1D solutions. We observe that the Darcy-based model aligns slightly better with the first half of the breakthrough curve, while the superficial velocity model aligns slightly better with the second half. 

The observed trends suggest that additional simulations with larger $N$ would likely cause the computational results to shift slightly to the left, while the 1D Darcy-based model would shift further to the right (due to the decreasing values of $u$). This would likely result in even closer agreement between the computational solutions and the homogenised 1D model than what is observed in the present study. This highlights a possible direction for future work: investigating convergence for finer microstructures (i.e., larger $N$) than those currently considered. However, such investigations would require more powerful computational resources than those available for the present study.

\section{Conclusions}
\label{sec:conc}

In this work, we have investigated the process of contaminant capture via column adsorption, which is one of the most viable methods to actively remove greenhouse gases from the atmosphere to combat global warming. We have formulated a 3D mathematical model that consists of an advection-diffusion equation for the contaminant transport, an equation describing the fluid motion for laminar incompressible flows, and an equation describing the adsorption dynamics of the contaminant on the walls of the porous media. A novelty of our approach is the description of adsorption through an equation defined on the surface of the porous medium, which is coupled to the advection-diffusion equation for the contaminant transport. In contrast, previous 3D models typically represented adsorption using simplified or effective boundary conditions for the advection-diffusion equation. 

We performed computational experiments involving the numerical solution of the 3D model on idealized microstructure configurations formed by spherical packings. Increasing the number of particles in each experiment, and adjusting the particles'  radius to keep the porosity constant, allowed to specifically investigate the effect of microstructure changes in the contaminant capture process. The analysis of concentration profiles and breakthrough curves of the different experiments showed that the microstructure configuration has minimal impact on the process, as the results are nearly indistinguishable. Of course, this conclusion is restricted to the level of approximations used in the current study, i.e. transport of laminar incompressible diluted mixtures in idealised porous media structures. 

Assuming a porous media formed by a periodic array of spheres and applying the method of multiple-scale homogenization, we rigorously derived an averaged 3D model of the process, which we then reduced to 1D. The resulting 1D model has the equivalent mathematical form than other 1D models used to describe column adsorption in the literature. However, our rigorous derivation yields a dispersion coefficient that explicitly incorporates microstructural details, whereas other models rely on experimental correlations or fitting procedures. 
The homogenization of the flow equations reduces the flow problem to Darcy's law. The assumed periodic array of spheres further enables the derivation of an explicit expression for the permeability, which depends on the porosity and scales with the particle's diameter, consistent with other popular effective permeability models, such as the well-known Kozeny-Carman equation.

To assess the ability of the 1D model to describe the column adsorption process, we first compared its numerical solution with results from 3D computational experiments using a slightly different arrangement of spheres designed to accommodate the low porosities typical of real experiments. Despite the difference in microstructure from that assumed in the homogenization, the agreement was remarkably good, with only minor discrepancies, particularly in the case with the fewest particles ($N=10$). To provide a more rigorous validation, we then performed 3D simulations using a periodic microstructure identical to that in the homogenization. By refining the microstructure while keeping the porosity constant, we demonstrated that the 3D solutions converge toward the 1D model predictions. These results confirm that the 1D model accurately captures the essential features of the column adsorption process, with minimal sensitivity to microstructural variations.

%In order to assess the capacity of the 1D model to describe the column adsorption process, we solved the 1D model numerically and compared the resulting breakthrough curves and concentration profiles with those from the computational experiments. We found that the solution of the 1D model is extremely close to the results of the computational experiments, showing a slightly greater divergence with the experiment containing the smallest number of particles ($N=10$). These results confirm that 1D models can accurately describe column adsorption experiments, with minimal influence from the porous media microstructure, as long as the contaminant is sufficiently diluted in the mixture and flow rates in the column are slow. 

\section*{Author contributions}
MA, FF: Conceptualization, Methodology, Investigation, Formal analysis, Writing – original draft,  Writing – review \& editing. 

\section*{Acknowledgments}

This publication is part of the research project TED2021-131455A-I00 of the Agencia Estatal de Investigación (Spain), funding both authors. This publication is also part of the research projects PID2023-146332OB-C22 (funding M. Aguareles) and PID2023-146332OB-C21 (funding F. Font)  financed by the Spanish \emph{Agencia Estatal de Investigación} of \emph{Ministerio de Ciencia, Innovaci\'on y Universidades}. Also, M. Aguareles gratefully acknowledges the support of the “consolidated research group” (Ref. 2021-SGR-01352) of the Catalan Ministry of Research and Universities. F. Font gratefully acknowledges the SRG programme (Ref. 2021-SGR-01045) and the Serra-Hunter Programme of the Generalitat de Catalunya.

\section*{Conflict of interest}
All authors declare no conflicts of interest in this paper.

\end{document}